# What would it cost to connect the unconnected? Estimating global universal broadband infrastructure investment

Edward J. Oughton[12*], David Amaglobeli[3], and Marian Moszoro[3]

[1]Geography and Geoinformation Sciences, George Mason University, Fairfax, VA, 22030.
[2]Environmental Change Institute, University of Oxford, Oxford, Oxfordshire, UK, OX1 3QY.
[3]Fiscal Affairs Department, International Monetary Fund, Washington DC, 20431.

*Corresponding author: Edward Oughton, Exploratory Hall, 10431 Rivanna River Way, Fairfax, VA, 22030. Tel: 703-993-1210. Email: eoughton@gmu.edu.

**Abstract**

Roughly 3 billion citizens remain offline, equating to approximately 40 percent of the global population. Therefore, providing Internet connectivity is an essential part of the Sustainable Development Goals (SDGs) (Goal 9). In this paper a high-resolution global model is developed to evaluate the necessary investment requirements to achieve affordable universal broadband. The results indicate that approximately $418 billion needs to be mobilized to connect all unconnected citizens globally (targeting 40-50 GB/Month per user with 95 percent reliability). The bulk of additional investment is for emerging market economies (73 percent) and low-income developing countries (24 percent). To our knowledge, the paper contributes the first high-resolution global assessment which quantifies universal broadband investment at the sub-national level to achieve SDG Goal 9.

## I. Introduction

Broadband infrastructure consists of all electronic and non-electronic assets operated to provide broadband services to users. Indeed, investing in universal broadband infrastructure is essential for Internet adoption and is a key part of the United Nations Sustainable Development Goals via Goal 9 (United Nations, 2019). The availability of Internet connectivity can provide new economic opportunities for unconnected communities and help in fostering greater structural shifts in labor markets towards more productive, digitally-enabled activities (Jung and López-Bazo, 2020). Moreover, Internet-enabled online services are increasingly essential for economic development, with the literature documenting improved educational outcomes, enhanced healthcare, economic productivity benefits, and job creation (Briglauer and Palan, 2023; Czernich et al., 2011; Ford and Koutsky, 2005; Gallardo et al., 2020; Koutroumpis, 2019; Prieger, 2013a; Tranos and Mack, 2015; Whitacre et al., 2014).

Over the past four decades telecommunication liberalization has introduced new entrants to markets across the globe, curbed excessive pricing and helped to drive the adoption of new waves of telecom consumer technologies and services through competition (for example, from fixed telephones through to Internet-connected smartphones) (Abrardi and Cambini, 2019; Dunnewijk and Hultén, 2007; Genakos et al., 2018; Kongaut and Bohlin, 2014; Oughton et al., 2016). However, the coverage and adoption of these services is far from ubiquitous. Indeed, while the large disparity in broadband connectivity is a longstanding problem caused by many factors (e.g., availability, affordability, relevance, skills etc.) (Chinn and Fairlie, 2007), this issue has been made highly evident in recent years (such as during the COVID-19 pandemic). Many citizens across the global have shifted to utilizing online services during this time (Baarsma and Groenewegen, 2021; Chang and Meyerhoefer, 2021; Szász et al., 2022), leaving behind those not readily using the Internet. Only approximately 60 percent of the global population is online (World Bank, 2022).

Many of those citizens yet to be successfully connected to a decent broadband connection are in emerging market economies (2 billion) or low-income developing countries (1 billion) (del Portillo et al., 2021). For example, in parts of Emerging and Developing Asia (1.7 billion) or Sub-Saharan Africa (0.75 billion) (Alper and Miktus, 2019a; World Bank, 2022). During COVID-19 this disparity, known as the 'digital divide', prevented the implementation of effective social distancing policies as unconnected citizens could not work remotely. On the other hand, more digitalized industries experienced lower revenue losses (Abidi et al., 2022; Copestake et al., 2022). Also, areas with superior broadband exhibited better labor market resilience, saw improved educational outcomes, and citizens in these locations could more easily access government support (Isley and Low, 2022; Mac Domhnaill et al., 2021; Sun, 2021).

Placing broadband infrastructure at the center of economic policy agendas can unlock economic opportunities, create jobs, and generate growth as well as improve the quality of life of citizens (Briglauer et al., 2021; Prado and Bauer, 2021; Simione and Li, 2021; Strover et al., 2021). In addition, the provision and adoption of digital services which use broadband enables workers affected by technology-driven economic transformation to reskill while also ensuring households have access to basic services such as healthcare, education, the social safety net, and financial services (Khera et al., 2022; Lobo et al., 2020). Access to broadband also helps reduce depopulation in rural and remote areas (Lehtonen, 2020). Moreover, higher levels of digitalization expands the tax base and strengthens revenue collection (Falk and Hagsten, 2021), as well as transforming public financial management by modernizing relevant systems, improving public service delivery, enabling digital payments, and promoting transparency (Gupta et al., 2017). Wider broadband access can entail environmental benefits, as digitally connected economies can contribute to lower carbon emissions by leveraging connected technologies for 'smart' management of the energy sector, utilities, manufacturing processes, agriculture and land use, buildings, services, transportation, and traffic management (Lebrusán and Toutouh, 2020).

While investing in broadband infrastructure is key, by itself, it does not guarantee digital adoption (Manlove and Whitacre, 2019; Rosston and Wallsten, 2020). New Internet users need to have the necessary skills to access relevant content (e.g., in meaningful spoken language). Unfortunately, some poorer, less-educated communities may not always trust or know how to use these key technologies or may not have any proof of identity to access online secured services (Hasbi and Dubus, 2020; Liu and Wang, 2021). Legal and regulatory weaknesses also should be addressed to allow countries to take full advantage of all the benefits, such as user protection, data privacy, and cybersecurity.

Against this background, this paper makes two important contributions to the literature. Firstly, the research provides answers to the following important questions: (i) how much investment is required in broadband infrastructure to achieve affordable universal connectivity (e.g., to achieve a 90% adoption target), and (ii) how does data consumption and Quality of Service (QoS) affect the necessary investment? And secondly, to answer these questions we also contribute a high-resolution geospatial method capable of estimating local universal broadband investment costs for each country, with the sub-national results being aggregated to allow for cross-country comparisons by income group and region. We generally prioritize the use of 4G cellular technology to provide affordable access to Internet (using satellite broadband only as a last resort) and find that additional investment needs to provide universal broadband connectivity amounts to $418 billion globally or approximately 0.45 percent of global Gross Domestic Product (GDP) (in 2020 US$), when targeting 40-50 GB/Month per user with 95 percent reliability. In terms of investment as a share of GDP, the estimated additional spending needs are the largest in Sub-Saharan Africa (4.49 percent of GDP) and the smallest in advanced economies (0.02 percent of GDP).

The rest of the paper is organized as follows. Section II provides a literature review on aspects pertaining to broadband connectivity and then Section III describes the method for costing broadband infrastructure investments. Next, Section IV discusses the cost drivers of broadband infrastructure investment and Section V provides estimates of additional investment needs. Finally, Section VI provides a sensitivity analysis around the cost estimates, with conclusions being provided in Section VII.

## II. Literature Review

There are many economic benefits that result from enhanced broadband coverage and adoption leading to enhanced economic growth, employment, regional development, productivity and firm performance benefits (Bertschek et al., 2015; Hussain et al., 2021; Sawng et al., 2021; Stamopoulos et al., 2022; Vu et al., 2020). For example, the diffusion of mobile broadband has been evaluated for its impact on GDP for 135 countries between 2002–2014 concluding that on average a 10 percent increase in mobile broadband adoption causes a 0.8 percent increase in GDP (Edquist et al., 2018). Moreover, panel econometric analysis for 27 European Union member states between 2003 and 2015 finds a small but significant effect of full Fiber-To-The-Premises broadband adoption, whereby a 1 percent adoption increase has an incremental GDP gain of 0.002–0.005 percent (Briglauer and Gugler, 2019). There is also causal evidence to suggest that broadband infrastructure and adoption positively affect firm size when captured by either sales or employment (DeStefano et al., 2018; Karim et al., 2022). These benefits can be influenced by the income level of the country in question, as low income countries can derive significantly higher gains from mobile broadband due to the lack of existing fixed broadband infrastructure (Thompson and Garbacz, 2011). There are also local firm formation impacts, with a one standard deviation higher broadband download speed being associated with a 2.1 percentage point increase in the annual growth rate of businesses (Chen et al., 2022).

A wide range of broadband technologies exists for users to connect to the Internet, each with different cost profiles (Schneir et al., 2023). Firstly, fixed technologies utilize a physical fiber optic or legacy copper/coaxial cable to provide broadband services to users. However, upfront capital deployment costs for Fiber-To-The-Premises can be prohibitive, reaching >$800 per urban premises and >$2,600 per rural premises (Rendon Schneir and Xiong, 2014), making this approach most common in advanced high-income countries where users are able to pay more (although operational costs once installed can be relatively low). In contrast, wireless technologies including mobile cellular (e.g., 4G) and satellite network architectures, provide lower-cost broadband services by eliminating the expense of laying a physical cable to each user (yet providing a lower QoS) (Osoro et al., 2023). For example, the total deployment and operational cost for mobile broadband can range from $300 per urban user to $1,500 per rural user for data packages offering ~30 GB/Month (Oughton, 2022). However, wireless methods still require a fixed high-capacity connection at some point in the data transmission process (e.g., fiber), so that information can be exchanged with the wider Internet. The focus of the assessment in this paper mainly focuses on using 4G as a low-cost technology for delivering wide-area mobile broadband.

The fourth generation of cellular technology has been hugely successful, with 4G Long Term Evolution (LTE) being deployed from approximately 2010 onwards, delivering wide-area mass market consumer broadband services (Lehr et al., 2021). The success of 4G was propelled by consumer desire to purchase novel smartphone devices, instigated by the success of Apple's iPhone (West and Mace, 2010). While 3G provided a very basic data rate to users (e.g., below a peak of 10 Mbps, and often experienced below 0.5 Mbps per user), the deployment of 4G meant that users could stream video content over their mobile devices, at speeds up to 100 Mbps, with a regular experienced speed of 2-10 Mbps. More recently, 5G has introduced new capabilities that were not possible with 4G, including enhanced mobile broadband and ultra-reliable low-latency communication (Cave, 2018a; Oughton et al., 2021; Rendon et al., 2021, 2019). For many developing countries, the technology has several limitations which make it unsuitable for providing universal broadband (Forge and Vu, 2020). In recognition of this, recent 5G developments have focused on new Reduced Capability devices (RedCap) to reduce cost (Veedu et al., 2022), however this will still take time to come to fruition.

Despite important gains over recent years in connecting unconnected Internet users, access to mobile broadband is far from universal. Most regions of the world have high 4G population coverage, except Sub-Saharan Africa and the Middle East, where 2G voice and basic 3G data access can be more prevalent (ITU, 2022). There are multiple reasons for the digital divide and lack of adoption, including infrastructure availability, affordability, market structure, and a lack of human capital (Lythreatis et al., 2022). Certainly, many more infrastructure assets (e.g., sites, fiber etc.) need to be built to provide the required coverage necessary to serve higher adoption levels. These sites largely need building in coverage-constrained rural and remote areas (where no existing assets exist to provide broadband services), rather than in capacity-constrained areas (such as urban and suburban areas where capacity limits have been reached).

Within countries, there are also significant differences between urban, suburban, and rural settlements. The heterogeneity in access to the Internet and the availability of a computer is also quite considerable. Often technologies can be slow to reach ubiquity as they follow a logistic 's-shaped' adoption curve, with late adopters affected by a variety of barriers (Ramírez-Hassan and Carvajal-Rendón, 2021; Salemink et al., 2017; Whitacre, 2008). For example, adoption can be affected by challenges in supply-side infrastructure delivery and the stickiness of demand-side wages in low-income groups. In other words, the deployment of infrastructure gets harder and more complex, in economic viability terms, to serve the final unconnected population deciles.

The Alliance for Affordable Internet initially set an affordability threshold at whether 1 GB of mobile broadband data is priced at 2% or less of the average monthly income in a country. The UN Broadband Commission later adopted this "1 for 2" target as a standard for affordable Internet. More recently, this has been updated to 5 GB priced at less than 2% or less of the average monthly income in a country (Alliance for Affordable Internet, 2022). However, while 88% of the world's population live within areas covered by at least one 4G network (ITU, 2022), approximately one third of citizens remain offline due to the high cost of Internet access relative to income levels.

Data consumption per user varies greatly depending on the market. For example, in 2022 a mean global smartphone user consumed an average of 15 GB per month, but this is as low as 5 GB per month in Sub-Saharan Africa, or as high as 25 GB per month in India or parts of Southwest Asia (e.g., the UAE). Moreover, by 2028 this is forecast to reach 46 GB per month for a mean global smartphone user, compared to a low of 18 GB per month in Sub-Saharan Africa, and a high of 55 GB per month in North America and North East Asia (Ericsson, 2022).

### III. Method for Estimating the Cost of Broadband Infrastructure Investments

We develop an assessment model with an associated tool to estimate broadband infrastructure needs—the Digital Infrastructure Costing Estimator (DICE) (Oughton, 2023). While universal broadband infrastructure cost modeling is an established field, most assessments are usually carried out at the national level for a single country (Lee et al., 2021). Although country-specific assessments are extremely important for providing highly detailed strategic information, a key weakness is that such approaches have poor external statistical generalization of the key findings to other contexts. Indeed, only recently have research efforts begun to take more of a cross-country global assessment approach (Oughton, 2022; Oughton and Lehr, 2022), utilizing harmonized global datasets for model inputs, to provide enhanced systematic insight across many markets. A key challenge in enabling this type of assessment has been developing a model, both general and flexible enough, to account for many different country contexts.

The DICE method enables the estimation of comparative country-specific investment in broadband infrastructure to achieve universal broadband connectivity. The model assumes investments in infrastructure to support predominantly terrestrial 4G deployment, while allowing for satellite connectivity in areas which require remote coverage in very hard-to-serve locations. The motivation for using wireless cellular connectivity is that this is well known to be one of the cheapest ways to provide wide-area broadband services affordably, especially for rural areas (Kumar and Oughton, 2023; Prieger, 2013b; Rendon Schneir and Xiong, 2016). Broadband connectivity is composed of the first mile (where and how connectivity enters the country), the middle mile (how data packets are transported long distances between different regions), and the last mile (how data packets are distributed locally to end-users). The method presented here estimates the total investment necessary to provide the infrastructure to connect currently unconnected citizens. It is also acknowledged that investments need to be made in fostering basic digital skills and content for end users and designing effective policies and broadband regulations to achieve universal broadband access. Effective policies could specifically target, for example, reducing prices via increasing market competition, or improving digital adoption for enabling financial inclusion.

Before quantifying the cost of building the necessary infrastructure, the model starts by estimating future data demand (Figure 1). The demand is estimated to obtain the required quantity of traffic to be served, followed by a network dimensioning process, and finally, the cost estimation metrics. Each of the three main model modules will now be presented in generalizable mathematical form (Demand, Network Dimensioning, and Costs), before the model data is explicitly outlined.

Importantly, the types of techno-economic models utilized by telecommunication policy decision makers for assessing broadband network investment are typically designed around 'geotypes' (essentially groups of similar areas placed into single market segments with common cost properties e.g., dense urban, urban, suburban, rural, remote etc.). The DICE model works by grouping local statistical areas into deciles, with the first decile containing 10% of the areas with the highest population density (broadly equivalent to dense urban), and the final decile containing 10% of the areas with the lowest population density (broadly equivalent to remote).

Figure 1. Sequential Method Illustrated as a Box Diagram

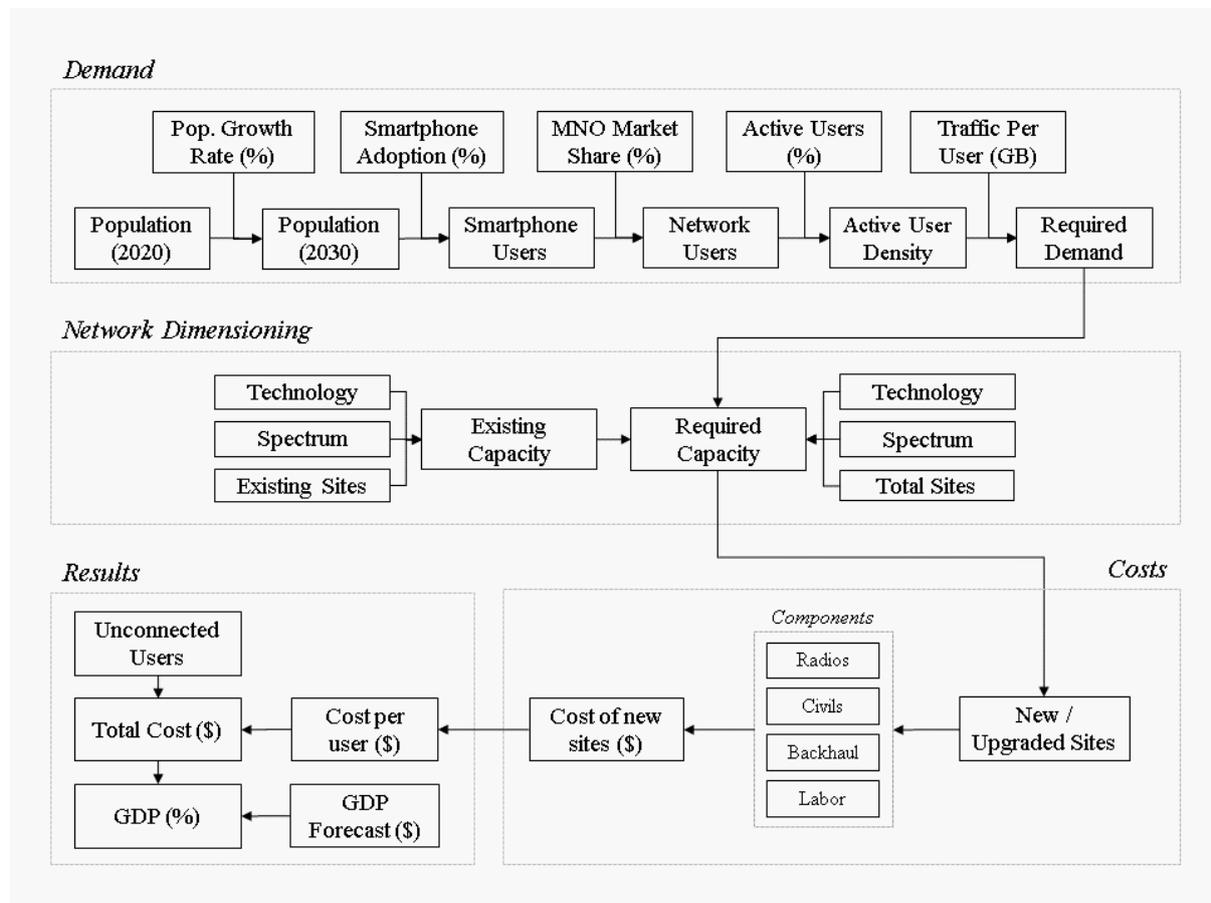

The analysis follows the country groupings utilized in the International Monetary Fund's World Economic Outlook which includes Advanced Economies (AEs), Emerging Market Economies (EMEs) and Low-Income Developing Countries (LIDCs) (IMF, 2023). This aim is to provide a meaningful way to group countries with general characteristics. These groupings are categorized based on per capita income, export diversification and the degree of global financial integration. Over time these groups have remained relatively static, with transition between groups a rarity (and generally only having taken place in recent times when countries have joined the Euro area, such as the accession of Lithuania).

**A. Estimation of Data Demand Density**

Estimating the amount of future data traffic requires obtaining the number of active smartphone users ($U_{it}$) in the $i^{th}$ population density decile for the busiest hour of the day ($t$), for all $n$ deciles. Using the current population in the starting year ($P_i$), the population growth rate ($G$), and the total number of years the assessment spans ($y$), the population in the final year (2030) that needs to be served is estimated. The final year ensures that the whole target population can be covered. However, it is still unlikely that 100 percent of the population will have adopted broadband even by 2030. Therefore, we set the expected smartphone adoption rate ($A_t$) for the end of the assessment period at 90 percent, which is still highly ambitious, but more realistic than 100 percent.

Next, as is common in telecommunications regulatory assessment, a 'hypothetical mobile network operator' is modeled and defined in terms of its user base (Falkner et al., 2000; Ioannou et al., 2021; Ofcom, 2018; Oughton et al., 2019). The hypothetical network operator serves a specific market share ($MS_t$) for the total population (e.g., 25 percent if there are four leading mobile operators), and has an average quantity of existing infrastructure assets, such as sites and spectrum (with the market share parameterized for each country using empirical OpenCelliD data). This step is necessary because we must model the traffic load for a single broadly representative network, to capture a broad balance between fixed and variable costs. This then enables the cost of serving a single user to be obtained. As the final step in the analysis, the cost of universal service will be calculated by multiplying the quantity of unconnected users by the cost of serving a single additional user.

Importantly, not all users access the network simultaneously. In fact, at any one time, only 5-20 percent of the customer base of a network may be exchanging traffic with a local site. Traffic is often generated in 'bursts' when a device downloads content, for example, adding chunks of video to a memory cache. Therefore, the percentage of active users ($AU_t$) exchanging traffic in a single moment needs to be set. Equation (1) specifies how we obtain the desired quantity of users which we need to model, as follows:

$$U_{it} = (P_i \cdot \left(1 + \left(\frac{G}{100}\right)\right)^y) \cdot \left(\frac{A_t}{100}\right) \cdot \left(\frac{MS_t}{100}\right) \cdot \left(\frac{AU_t}{100}\right) \quad (1)$$

To estimate the required data rate ($RD_t$) per user, a specified monthly data consumption quantity ($MD_t$) needs to be converted to a mean capacity in the busy hour of the day, as detailed in Equation (2).

$$RD_t = MD_t \cdot 1000 \cdot 8 \cdot \left(\frac{1}{n_d}\right) \cdot \left(\frac{f_{dh}}{100}\right) \cdot \left(\frac{1}{3600}\right) \qquad (2)$$

The data consumption per user ($MD_t$) is exogenously stated in Gigabytes based on the mean expected monthly consumption in the final year (2030) that needs to be served (e.g., 50 GB). This value is converted from Gigabytes to Megabytes and then from bytes to bits. Next, the monthly number of bits is converted into the daily demand based on the number of days per month ($n_d$). From the daily quantity of bits and the percentage of traffic which takes place in the busiest hour of the day ($f_{dh}$), the mean number of bits can be estimated for the hypothetical network's busiest daily hour. The mean hourly rate can then be converted to the *bits per second* (bps) rate based on the number of seconds in an hour (i.e., 3600).

Next, we quantify the demand density. Equation (3) describes how the number of active users ($U_{it}$), the required per-user data rate ($RD_t$), and the total geographic area of the decile ($Area_i$) can produce an estimate of the traffic demand density ($D_t$):

$$D_t = \frac{(U_{it} \cdot RD_t)}{Area_i} \qquad (3)$$

Once the demand density figure has been obtained, it is necessary to estimate (i) the network capacity based on the existing number of cell sites and (ii) the necessary number of new sites required to meet the demand density. The following section details the approach.

**B. Estimation of Necessary Infrastructure Upgrades**

The upgrade strategy adopted is to build greenfield 4G cellular towers using either a wireless or fiber backhaul to transport data traffic, either to another existing tower with fiber, or a newly built tower with newly built backhaul capacity. Where past 2G or 3G towers are present, costs can be saved by upgrading these to 4G Radio Access Network (RAN) equipment.

The capacity of a wireless cellular network depends on a combination of the density of cellular sites, the efficiency of the technology (e.g., 4G), and the bandwidth of the spectrum channels available to exchange information over (e.g., 10 MHz) (Frias et al., 2020; Oughton et al., 2020). To calculate the area capacity ($C_{it}$) (Mbps/km²) of the $i^{th}$ decile at time $t$, it is necessary for each spectrum frequency ($f$) to multiply the spectral efficiency ($\eta_{Area}^{f}$) by the available channel bandwidth ($BW^{f}$). Then all available frequencies must be summed, as per Equation (4):

$$C_{it} = \sum_{f} \eta_{Area}^{f} BW^{f} \qquad (4)$$

The available bandwidth ($BW^{f}$) can be selected for a country based on the current average spectrum portfolio available to the hypothetical network operator being modeled. Generally, low-income countries have smaller spectrum portfolios than emerging economies or advanced economies, as demand is lower in the former countries. The mean network spectral efficiency ($\bar{\eta}_{Area}^{f}$) for each frequency must be estimated stochastically via Monte Carlo simulation. Based on the average number of cells per site ($\bar{\eta}_{cells}^{f}$) (e.g., three) and the density of cellular sites transmitting in the same frequency ($\bar{\eta}_{sites}^{f}$), this can be estimated as per Equation (5).

$$\bar{\eta}_{Area}^{f} = \bar{\eta}_{cells}^{f} \cdot \bar{\eta}_{sites}^{f} \qquad (5)$$

The spectral efficiency is a function of the signal-to-interference-to-noise ratio (SINR), which is affected by the distance from the cell site and interference from other radio transmitters. Using network simulations, it is possible to generate performance curves based on the probability density function for the capacity achievable with different cell sizes and, thus, different site densities (Mogensen et al., 2007). For a desired QoS level (e.g., 95 percent reliability), the capacity can be estimated for each site density using Monte Carlo simulations based on the free space path loss, with log-normal shadow fading ($\mu = 2, \sigma = 10$). Thus, lookup tables can be generated from these simulations, following an established method in the literature (Oughton and Jha, 2021). Using these lookup tables, it is then possible to estimate (i) the current capacity of an area based on the existing site density ($ExistingSites$) and (ii) the total number of sites ($TotalSites$) to provide a network capacity that meets the demand density value ($D_t$). Therefore, the number of new or upgraded sites that need to be built ($NewSites$) can be estimated as per Equation (6):

$$NewSites_i = TotalSites_i - ExistingSites_i \qquad (6)$$

As the number of required sites can now be obtained, the next step is to estimate the required cost of building and operating this infrastructure until the end of this decade.

**C. Estimation of Costs**

The total cost of new sites can be estimated for each $i^{th}$ decile based on the necessary *Capex* ($c$) and *Opex* ($o$) investment needed in the RAN ($RAN_i$), the backhaul cost to connect each site ($Backhaul_i$) into a fiber Internet Point of Presence (PoP), the civil engineering materials cost of building a cellular tower ($Civils_i$), and either a grid-connected power system or one capable of generating renewable energy (solar or wind), and storing it via a battery pack ($PowerSystem_i$). Finally, the labor cost is included for radio network planning ($Labor\_Planning_i$), equipment transportation ($Labor\_Logistics_i$), tower construction ($Labor\_Construction_i$), and equipment installation at the cellular site ($Labor\_Installation_i$). Equation (7) specifies how a new site's total capex ($Capex_i$) is estimated.

$$Capex_i = RAN_{ic} + Backhaul_{ic} + Civils_{ic} + PowerSystem_{ic} + Labor\_Planning_{ic}$$
$$+ Labor\_Logistics_{ic} + Labor\_Construction_{ic} + Labor\_Installation_{ic} \quad (7)$$

New sites built need to be connected to a wider network to exchange data traffic with the Internet, thus requiring necessary fiber connections to connect each site ($Metro\_Core\_Fiber_i$). In Equation (8), the mean distance between all new and existing sites ($TotalSites_i$) with fiber backhaul is first estimated. The mean distance is then multiplied by the number of new sites, a backhaul-core splitting factor ($\alpha$, e.g., 10 percent), and the cost per fiber meter ($Fiber\_Cost_i$).

$$Metro\_Core\_Fiber_i = \left(\sqrt{\frac{1}{TotalSites_i}}\Big/2\right) \cdot NewSites_i \cdot \alpha \cdot Fiber\_Cost_i \quad (8)$$

The cost to upgrade an existing site is treated as the same as Equation (7), except the civil engineering cost ($Civils_i$) of building a new tower is excluded. This is justified because upgrading a 2G/3G tower to 4G would require new RAN equipment and most likely upgrading an old wireless microwave backhaul to a newer version with higher spectral efficiency (as well as adding a new power system, along with the necessary labor time for the overall upgrade).

Next, we estimate operational expenditures ($Opex_i$) for the network based on future annual operational expenditure up to 2030. The $Opex$ ($o$) is estimated to be approximately 15 percent of the initial asset value annually, plus any labor costs associated. This accounts for the ongoing cost of servicing and maintaining terrestrial 4G cellular sites providing mobile broadband. Equation (9) specifies the cost structure.

$$Opex_i = RAN_{io} + Backhaul_{io} + Civils_{io} + PowerSystem_{io} + Labor\_Planning_{io}$$
$$+ Labor\_Logistics_{io} + Labor\_Installation_{io} \quad (9)$$

For all new users connected, an additional cost is allocated to two key categories of necessary spending, including policy and regulation and online skills and content, based on the ITU Connecting Humanity assessment (International Telecommunication Union, 2020). Firstly, it is necessary to ensure resources are available for delivering high-quality governance by the telecommunication regulator ($Policy\_Regulation_i$, e.g., roughly \$2/user). Secondly, it is essential to ensure users have the necessary skills and available content to ensure Internet access is useful ($ICT\_Skills\_Content_i$, e.g., \$12/user); otherwise, this will be a crucial barrier to adoption (and, thus, attaining the key benefits associated with digital adoption).

In hard-to-reach regions, where the estimated costs of building terrestrial infrastructure exceed provision via satellite, a cost per user is allocated for broadband services provided from either Geosynchronous or Low Earth Orbit satellite constellations ($Remote\_Coverage_i$). Broadly following the ITU Connecting Humanity approach (International Telecommunication Union, 2020), this is based on a \$200 monthly subscription cost split between 12, 8 or 4 users, for LDICs, EMEs or AEs, respectively.

Next, the mean Total Cost of Ownership Per User ($TPU_c$) in each country ($c$) can be obtained by summing all relevant cost components, and the dividing by the total number of network users, as detailed below in Equation (10).

$$TPU_c = \frac{\sum_{i=1}^{n} Capex_i + Metro\_Core\_Fiber_i + Opex_i + Policy_i + Skills\_Content_i}{((\sum_{i=1}^{n} P_i \cdot (1 + (G/100)^y) \cdot \left(\frac{A_t}{100}\right) \cdot \left(\frac{MS_t}{100}\right)} \quad (10)$$

Finally, the Total Cost ($TC_c$) per country ($c$) can be obtained by multiplying the Total Cost of Ownership Per User ($TPU_c$) by the number of unconnected users ($UnconnectedUsers_c$) in each country ($c$), as stated in Equation (11).

$$TC_c = TPU_c \cdot UnconnectedUsers_c \tag{11}$$

This approach produces all necessary metrics for the model to support future investment decisions based on the standard or other user-defined inputs set. Having described the mathematical structure of the approach, the method can now focus on how the model will be populated with data and other parameter values.

**D. Model Data**

The DICE model needs populating with numerous different datasets, as summarized in Table 1. Firstly, the approach begins by estimating the population density for each local statistical area in each country. The global WorldPop 1 km$^2$ population mosaic is obtained for 2020 to provide insight into the local population spatial distribution (Tatem, 2017; WorldPop, 2019), and is intersected with either layer 1 or 2 boundaries from the Global Database of Administrative Areas (depending on the country) (GADM, 2019). Annual population growth forecasts for each country are then taken from the United Nations (IMF, 2021) to estimate the population density in 2030. Next, the local statistical areas for each country are grouped into deciles based on population density, so that the highest population density local statistical areas are allocated to the first decile and the lowest density areas are allocated to the bottom (tenth) decile. This allows network designs to be developed for each density decile (with cost related to user density).

Tower data estimates are obtained from TowerXchange annual reports to build insight into the number of existing accessible sites per country (TowerXchange, 2019a, 2019b, 2018a, 2018b, 2017). Where TowerXchange information is missing for a particular country, spatial data are taken from OpenCelliD, and the total sites are estimated based on a mean number of seven cells per site, as detailed on the OpenCelliD website FAQ (OpenCelliD, 2022). Non-4G sites are allocated to deciles based on an estimate of 2G sites per population coverage using ITU statistical data (ITU, 2022), with this allocation beginning in the densest deciles first. The same approach is used to estimate the number of 4G sites by using 4G population coverage obtained again from the ITU (ITU, 2022), with distribution beginning in the densest deciles. Once there are no more available sites to allocate from the total per country, the bottom deciles can end up with no sites to provide coverage, which matches the distribution seen (because operators rationally deploy infrastructure, building sites that serve present and future demand, which is higher in cities and lower in rural and remote areas).

**Table 1. Data Sources**

| Data | Unit | Spatial resolution | Source |
|---|---|---|---|
| Statistical boundaries | N/A | GADM levels 1 and 2 | (GADM, 2019) |
| Population count | Persons | 1 km$^2$ | (WorldPop, 2019) |
| Population growth rate | Percent | Country | (IMF, 2021) |
| Tower data | Sites | Country | (TowerXchange, 2019b, 2019a, 2019c, 2018a, 2018b, 2017) |
| Cell data | Cells | Coordinates | (OpenCelliD, 2022) |
| Cellular coverage | Percent | Country | (ITU, 2022) |
| Backhaul technology | Percent | Continent | (GSMA, 2019) |
| Labor costs | USD | Country | (IMF, 2022) |
| Internet adoption | Percent | Country | (World Bank, 2022) |

Next, the number of major network operators is derived from OpenCelliD data based on the aggregation of 2G, 3G, 4G, and 5G cells per country (OpenCelliD, 2022). In terms of the number of sites with existing fiber backhaul, GSMA estimates for 2025 are utilized as detailed geospatial information is not widely available (81 percent in North East Asia, 73 percent for North America, 33 percent for Europe, 21 percent for Latin America and the Caribbean, 20 percent for the Middle East and North Africa, 17 percent for South and South East Asia, and 15 percent for Sub-Saharan Africa) (GSMA, 2019). Data traffic of 50 GB/Month is targeted in EMEs and AEs, and 40 GB/Month in LIDCs, with 95% reliability. These are highly ambitious targets which reflect the upper bounds of more optimistic data traffic forecasts by cellular equipment vendors (Ericsson, 2022), for example, as Sub-Saharan Africa alone is only expected to reach 18 GB/Month in 2028.

Costs are derived per country based on the necessary work hours and the hourly cost of labor (IMF, 2022). For site planning & surveying, logistics, construction, and installation, 16 hours are allocated to each labor component. The estimated mean cost per hour for Information Communication Technology (ICT), logistics, and construction is developed for each country based on IMF calculations.[1] Finally, the labor cost is estimated by multiplying the mean number of hours by the mean hourly labor cost. For planning & surveying, and installation, the ICT labor cost is used, because this is specialist skilled work, whereas derived logistics and construction labor costs are directly used for these categories.

Finally, statistical data is taken from the World Bank open data portal on the percentage of individuals using the Internet (World Bank, 2022) to estimate the total number of unconnected users in each country. Next, the model context will be presented based on the collation of these data sets.

IV. Cost Drivers for Broadband Infrastructure Investment

The costs of delivering broadband infrastructure will be significantly determined by population density. Low population density areas are going to have poor infrastructure economics, as fixed investment costs must be split across a few potential users. Figure 2 visualizes the population density deciles for each country. Each decile represents 10 percent of the local statistical areas for each country in the sample, with the first decile (Decile 1) containing the top 10 percent of areas with the highest population density and the last decile (Decile 10) representing the bottom 10 percent of areas with the lowest population density. This approach emphasizes the areas of high population density across all populated continents. While also most notably illustrating the least populated areas in the hardest-to-reach deciles, spanning high-latitude regions in Canada, Russia, and Argentina, as well as the Amazon rainforest, the Sahara Desert, the Tibetan plateau, and much of Australia.

---

[1] Hourly rates by sector by country come from the International Labour Organization's statistics on wages (see https://ilostat.ilo.org/topics/wages/, accessed in September 2022). Missing values were imputed by (i) regressing the log of observed hourly wages on the log of GDP per capita, (ii) predicting the missing values with the observed log of the GDP per capita, and (iii) taking the exponential of the predicted values. The models' fitness for all sectors is above 0.9.

Figure 2. Visualizing Population Density Deciles for each IMF Country as the Sub-national Level Across 60,926 Local Statistical Areas

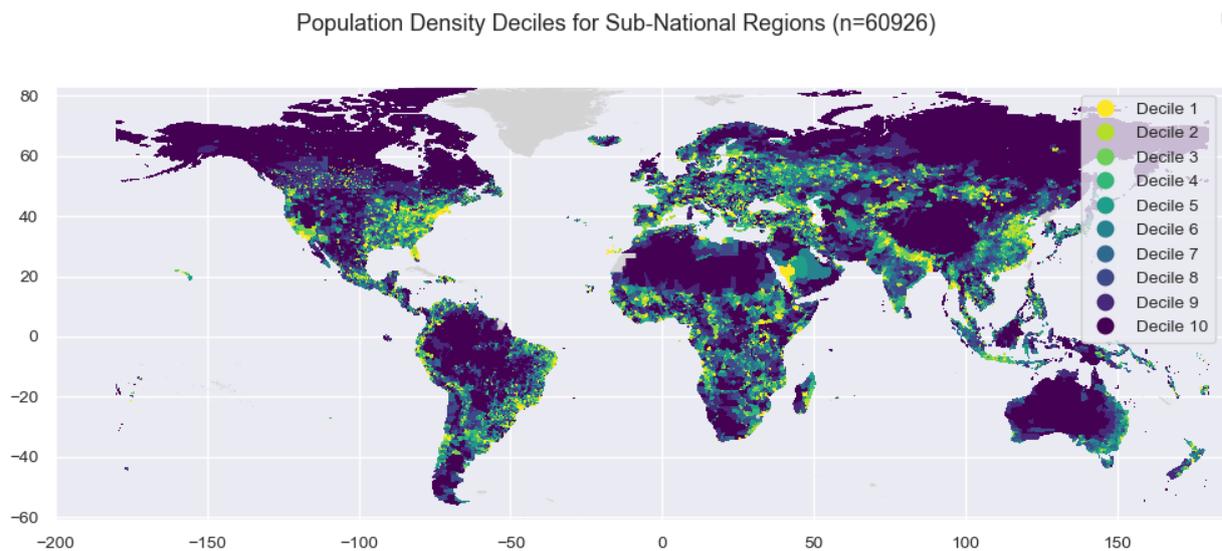

These deciles are visualized further in Figure 3 to illustrate the key dimensions of the infrastructure investment context for universal broadband. The following key observations can be made:

- The first decile logically has by far the largest population across all country income groups, as this segment contains most of the dense urban areas (Figure 3.A). However, the shape of the population distribution is different across income groups based on the level of underlying urbanization. For example, in advanced economies, where the urbanization rate is as high as 80 percent, there is a significant decline in the overall population after the first decile of 420 million people (39 percent of the total population) to as low as 22 million people in the final least-dense decile (2 percent of the total population). In contrast, in low-income developing countries, there is a less profound decrease in the population distribution across the deciles, with 317 million people in the densest decile (21 percent) and then 70 million in the final least-dense decile (5 percent).

- In terms of area, the first decile consists of 2 percent of the overall area, whereas the bottom decile consists of 51 percent, 45 percent, and 31 percent for advanced, emerging market, and low-income developing countries, respectively (Figure 3.B). The geographic area matters for infrastructure investments because the larger the area needing to be covered, the greater the magnitude of the required investment, given that there is more space to mediate, for example, in transporting data traffic to provide broadband services.

- In terms of population density, which combines the previous two metrics (population and area), emerging market economies have, on average, a higher density (Figure 3.C). For example, in the densest decile (decile 1), emerging market economies have an average population density of 5,901 persons per square kilometer compared with 659 and 4,640 persons per square kilometer in the advanced and low-income countries, respectively. For the least dense decile (decile 10), in advanced economies, the density falls to 1 person per square kilometer, which implies that even with higher consumer expenditure to spend on broadband services, it is challenging to deploy even a basic terrestrial broadband service.

Figure 3. Visualizing Key Country Metrics for Population Density Deciles by IMF Income Groups and Regions

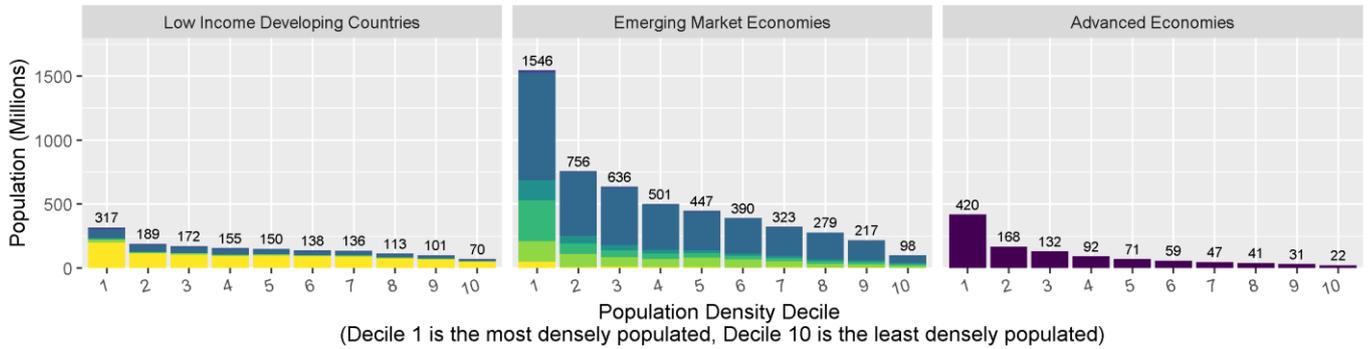

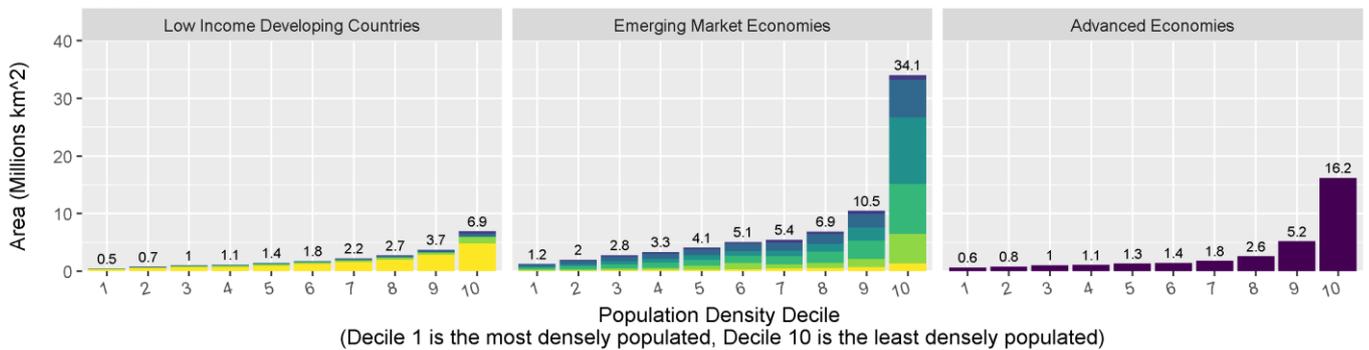

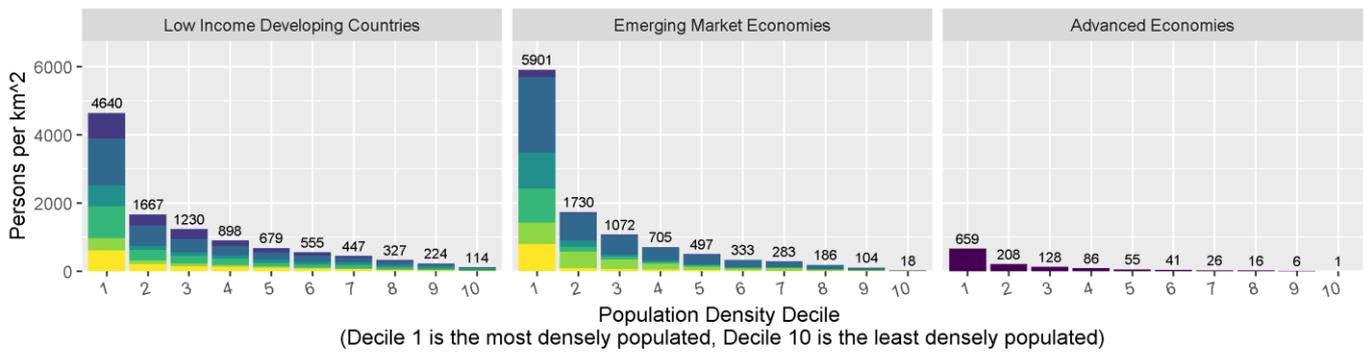

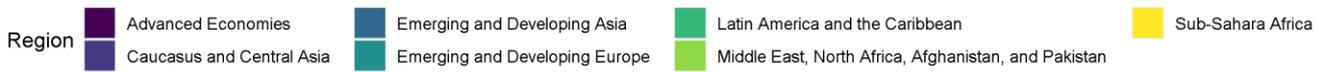

## V. Broadband Infrastructure Investment Needs

The aggregate global broadband infrastructure investment need is $418 billion.[2] This estimate is based on providing universal 4G cellular broadband to users with approximately 40-50 GB of monthly data at 95 percent reliability. To put this figure into context, annual MNO capital expenditure for 2023-2030 is estimated to be approximately $1.5 trillion ($187.5 billion annually) (GSMA Intelligence, 2023), meaning this investment is broadly equivalent to 2.3x the global annual capex of the mobile industry. Overall, the most prominent needs are in emerging market economies, estimated at $305 billion (73 percent), compared with a very modest investment in advanced economies of only $11 billion (3 percent) and a total of $102 billion (24 percent) in low-income developing countries (Figure 4.A). Indeed, the composition of investment required in low-income developing countries needs to be targeted into broadband infrastructure capital expenditure (27 percent), metro and backbone fiber (24 percent), and infrastructure operational expenditure (33 percent). This variance is explained by underlying differences in the existing infrastructure level (which reflects historical sunk cost investments) and population density differences.

In terms of geography (Figure 4.B), Emerging and Developing Asia has the largest needs at $176 billion (42 percent), followed by Sub-Saharan Africa ($91 billion, 22 percent), and the Middle East and North Africa, Afghanistan, and Pakistan ($69 billion, 17 percent). The regions requiring the least investment include Emerging and Developing Europe ($14 billion, 3 percent) and Caucasus and Central Asia ($7 billion, 2 percent), with these lower investment amounts related to the level of past investment or smaller populations compared to other regions (e.g., in the Caucasus). These investment quantities are heavily correlated with the number of currently unconnected users, which is the key driver of new broadband infrastructure.

---

[2] Compared with other infrastructure investment needs, the estimated cost of building digital infrastructure is much lower. For example, based on the IMF's previous estimates of additional spending required to meet infrastructure-related SDGs, the total investment required for electricity, roads, and water and sanitation are $4 trillion, $7.5 trillion, and $1.5 trillion, respectively.

Figure 4. Estimate of Required Investment by Income Group and Region

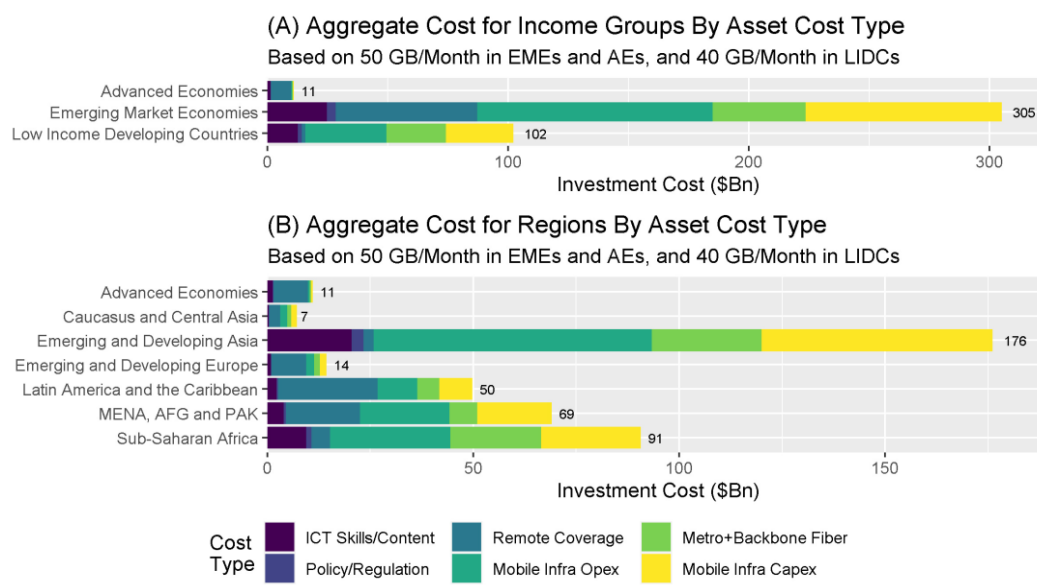

These aggregate costs mask the heterogeneity of investment needs across different population deciles, which is important in terms of digital inclusion and equity. For example, in emerging and developing economies, the distribution of investment by decile follows a bell-shaped curve (Figure 5.A), whereas in low-income developing countries, investment needs range from $3.6 billion in the first decile to $13.8 billion in the fifth decile, and down to $6.7 billion in the bottom decile. In contrast, advanced economies are estimated to only need modest investment in deciles five to ten, with investment below $2.5 billion in each decile category.[3] In terms of the regional composition (Figure 5.B), the largest investment needed across all deciles is visible in the distribution of Emerging and Developing Asia (up to $28.3 billion per decile), followed by Sub-Saharan Africa (up to $11.9 billion per decile), Latin America and the Caribbean (up to $9.3 billion per decile), and then MENA, Afghanistan, and Pakistan (up to $9.2 billion per decile). The lowest investment needs are estimated in Emerging and Developing Europe (up to $3.4 billion per decile) and the Caucasus and Central Asia regions (up to $1.7 billion per decile).

---

[3] One caveat to these estimates is that these will be prime locations for 5G deployment in the densest urban areas over the next decade. Thus, new spectrum and technology could support a proportion of this capacity if consumers can afford to purchase a 5G handset to access the network.

Figure 5. Country-level Estimates of Required Investment

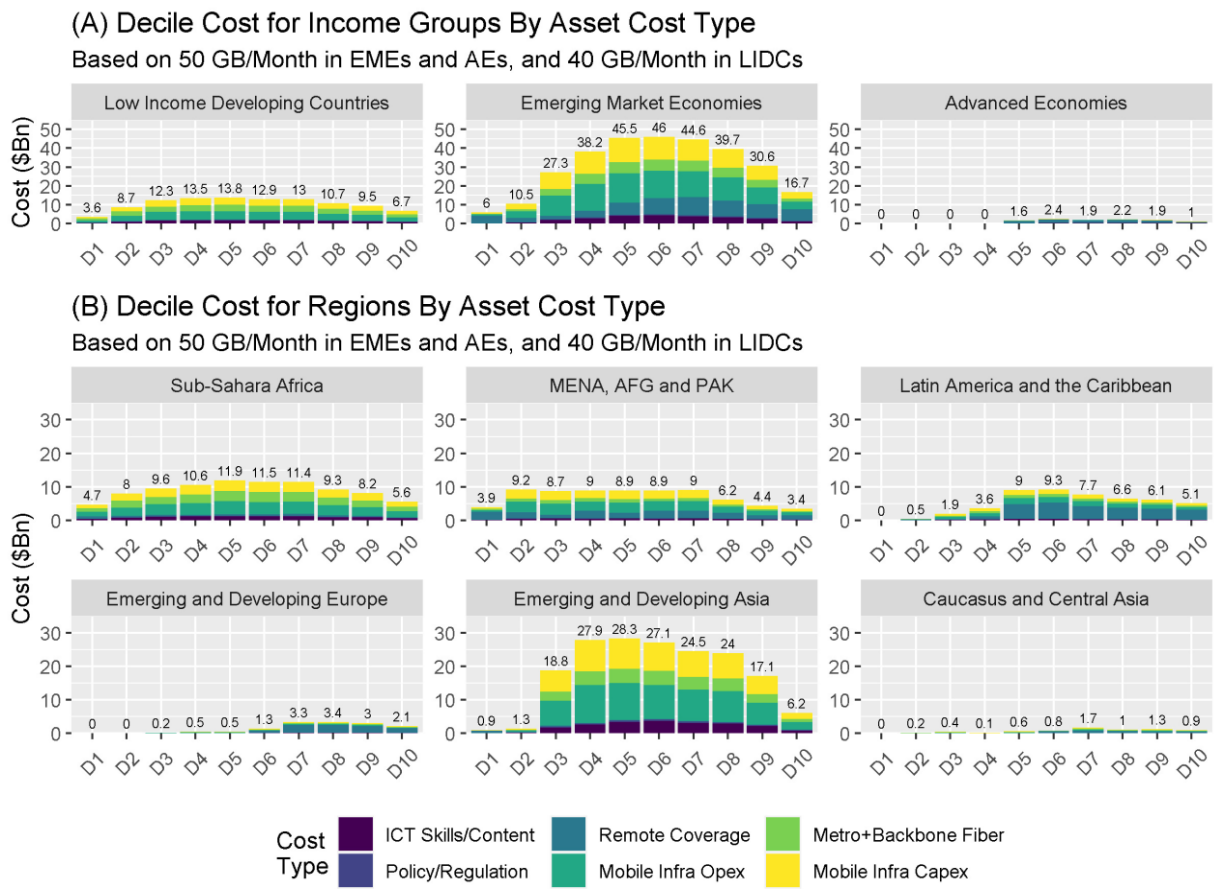

The picture considerably changes when looking at broadband infrastructure investment needs in percent of national GDP (in 2020 US$). The total investment costs in low-income developing countries are the largest at about 3.5 percent of GDP, followed by emerging market economies at 0.7 percent of GDP (Figure 6.A). Regarding geography, the largest needs are in Sub-Saharan Africa at 4.5 percent of GDP, followed by MENAP at 1.7 percent of GDP, Caucasus and Central Asia at 1.6 percent, and Latin America and the Caribbean at 1 percent of GDP (Figure 6.B).

Figure 6. Proportion of Necessary Investment Relative to GDP

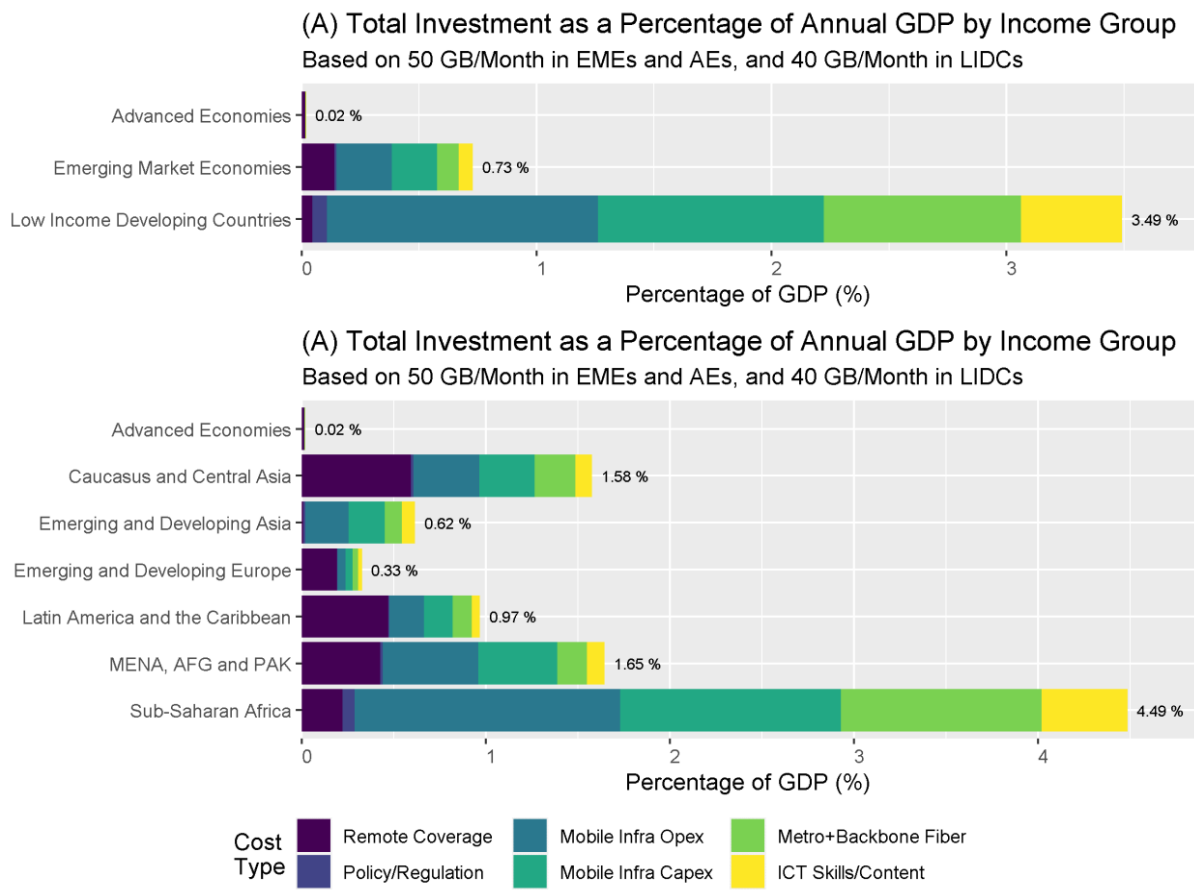

Since our modeling approach allows for the ability to provide a sub-national understanding of the costs of universal broadband connectivity, the granular spatial distribution of required investment can be presented (Figure 7). Areas requiring investment are colored on a scale of below $10 million (yellow) up to those requiring more than $100 million (dark purple). Unsurprisingly, those areas where the most investment is required are in high latitude regions (e.g., Northern Canada, Russia), the Sahara, the Amazon, much of central and western Australia, and the Tibetan Plateau. This results from these areas having many unconnected users and/or statistical boundaries covering larger geographic areas. Those areas requiring no investment are indicated in dark grey, which include the United States' coastal areas, Latin America, Western and Southern Europe, Eastern China, and large parts of India.

Figure 7. Visualizing Investment Costs for Population Density Deciles for each IMF Country at the Sub-national Level

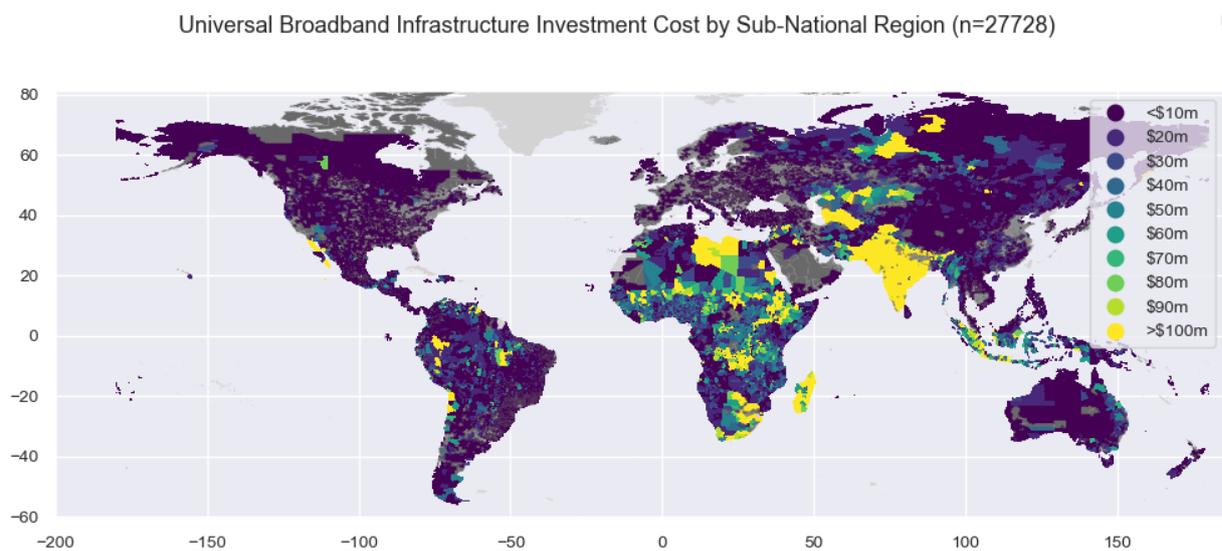

Note: Light grey indicates non-IMF countries. Dark grey indicates areas where no investment is required. The sample size of sub-national statistical areas is denoted by $n$.

The estimated universal broadband investment from the high-resolution methodology developed here is commensurate with other aggregated country-level estimates. For example, the ITU Connecting Humanity report (International Telecommunication Union, 2020), which is widely regarded as the key policy assessment appraising the cost of connecting unconnected citizens globally, estimated the necessary investment into broadband infrastructure to be around $428 billion, 1.4 percent higher than the estimate presented in this paper. Breaking down the comparison by spending category, geographic region, and country income groupings shows that (a) the largest difference is present in the cost of mobile infrastructure operational expenditure and (b) in the estimates for East Asia and the Pacific (Figure 8).[4] However, a key weakness of the ITU Connecting Humanity report is the lack of high-resolution sub-national estimates of investment.

---

[4] Unfortunately, without further details about the ITU's modeling approach, in terms of the data used and modeling assumptions (which are not available in the ITU report), it is not possible to further understand the causes of these divergences.

Another continental estimate by IMF staff reports the total investment needs of $14 billion for full 4G connectivity for Sub-Saharan Africa, which is significantly lower than the estimate presented in this paper. A key reason is the range of different methodological assumptions taken (Alper and Miktus, 2019b), including very high assumed spectral efficiencies of 15-23 bits per second per Hertz. Another large difference between the two estimates stems from conservative adoption and data consumption metrics, such as assuming a pre-2020 mean traffic rate per user in 2025 (thus, no data consumption growth). Indeed, with low monthly data consumption rates (less than 1 GB), this is insufficient to enable the range of necessary use cases targeted in this assessment, such as remote learning and video calling. Additionally, as carried out in this assessment, the evaluation only targets potential data demand in 2025, which is much more modest than targeting universal broadband in line with the SDGs. Finally, no topological network effects are accounted for, such as radio propagation and interference impacts, which can substantially impact the cost. Unfortunately, this continental analysis is isolated to Africa and does not provide a global assessment of the necessary universal broadband investment at the planetary scale.

Figure 8. Evaluation of the DICE Model Against the ITU Connecting Humanity Modeling Method

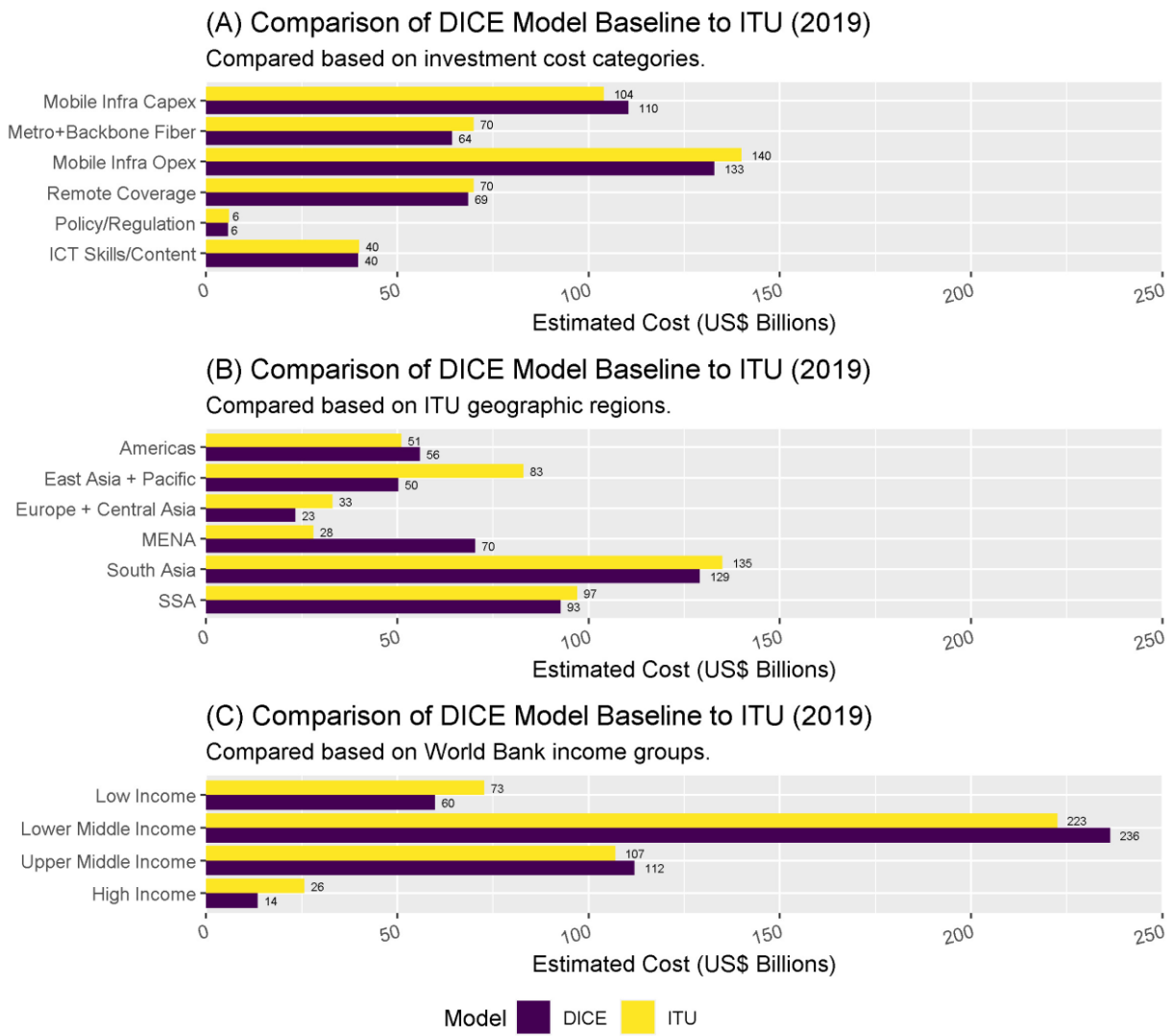

## VI. Sensitivity Analysis

The analysis indicates the high sensitivity of aggregate cost estimates to the data consumption target per user selected. Using a lower data consumption of 20 GB/Month per user in emerging market economies and advanced economies (as opposed to 50 GB under the baseline) and only 10 GB/Month per user in low-income developing countries (as opposed to 40 GB under the baseline), leads to a 52 percent decrease in the aggregate cost from $418 billion to $201 billion (Figure 9.A). This is because less infrastructure needs to be built to accommodate this lower data traffic target. One of the largest decreases is in emerging market economies, where the necessary investment decreases from $305

billion in the baseline to $136 billion in the reduced data consumption scenario (i.e., 55 percent reduction). On the other hand, the increase in the data consumption rate to 100 GB/Month per user in emerging market economies and advanced economies and 80 GB/Month in low-income developing countries results in an overall increase in estimated cost to $783 billion from $418 billion in the baseline (i.e., more than 90 percent rise). Again, the largest increase is in emerging market economies, with necessary investment escalating from $305 billion to $597 billion (growth of 49 percent), with low-income developing countries shifting from $102 billion to $172 billion (Figure 9.B).

Figure 9. Exploring the Impact of Data Consumption on Cost Sensitivity

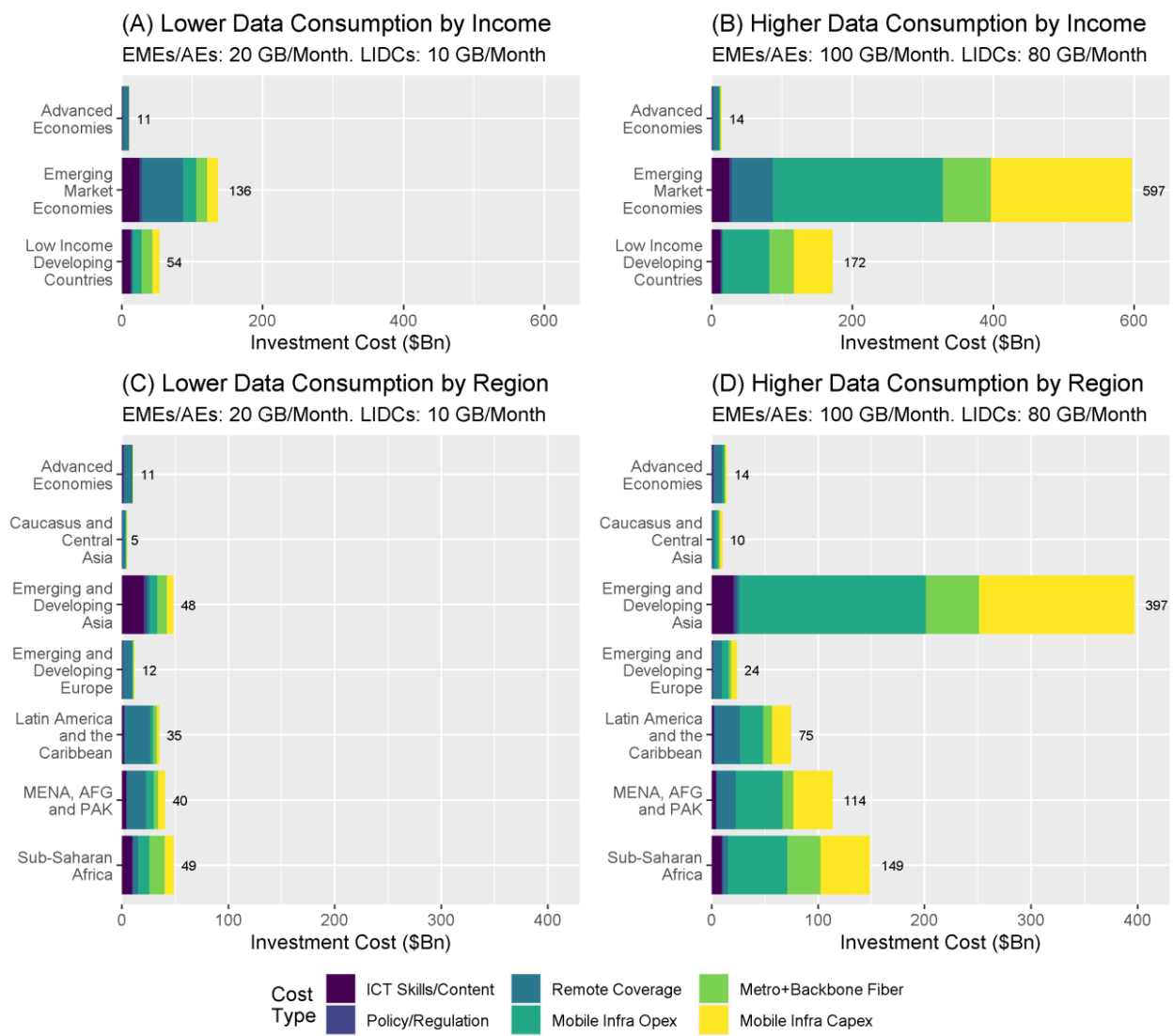

The estimated cost is also sensitive to varying the potential QoS for the provided broadband service. Compared to a 95 percent reliability level in the baseline, lowering service reliability to 5 percent (equivalent to an intermittent service in the busiest hour of the day) sees an overall cost decrease from $418 billion to $176 billion (Figure 10). There is a large decrease (77 percent) in emerging and developing Asia against the baseline, but a much more modest decrease (51 percent) in Sub-Saharan Africa, as considerable greenfield infrastructure still needs to be built even to deliver this reduced QoS level.

Figure 10. Exploring QoS Reliability on Cost Sensitivity

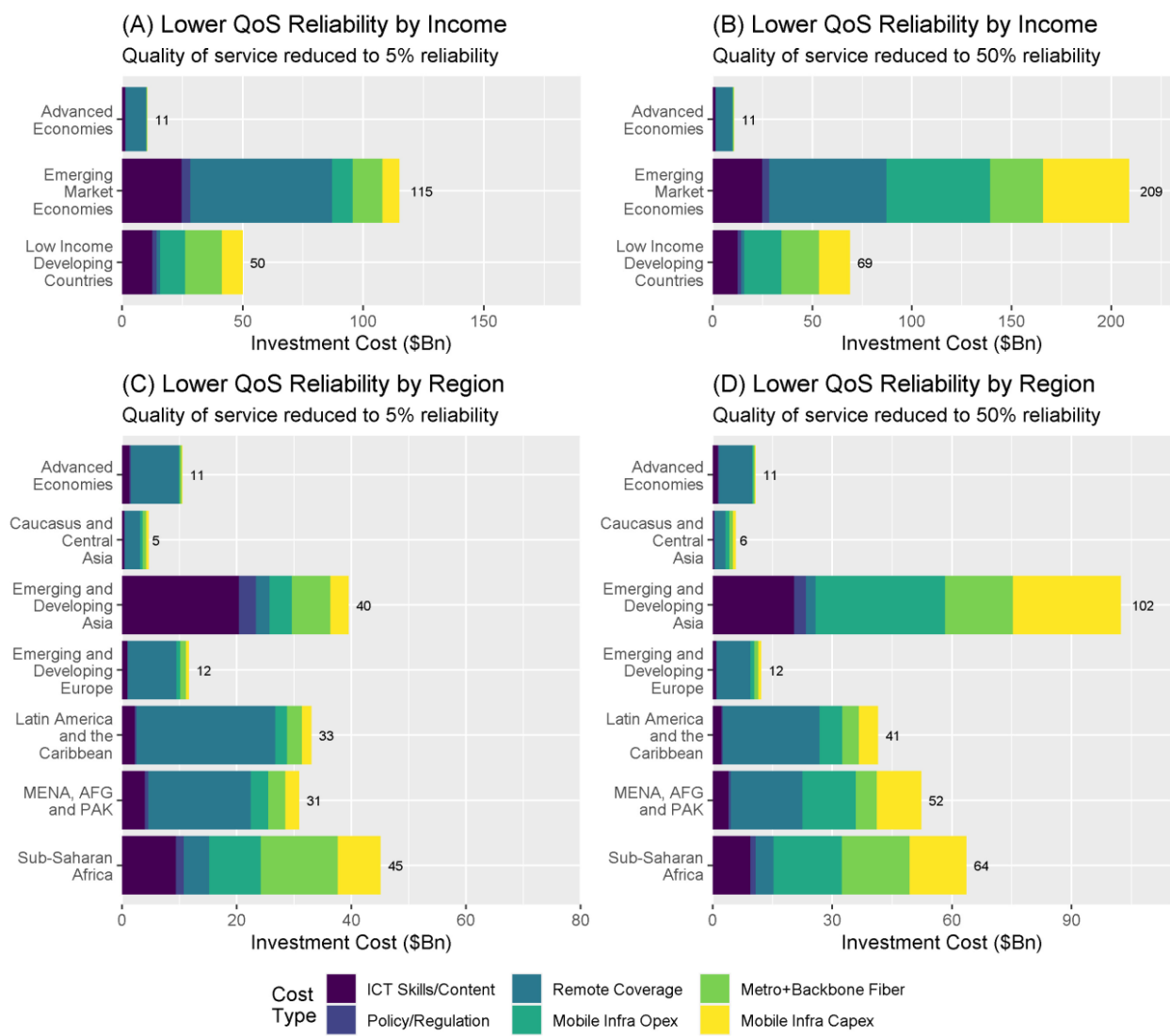

# VII. Conclusions

This paper assesses the cost of investing in broadband infrastructure to provide affordable universal broadband by 2030. To this end, we develop a new, open-source model—the Digital Infrastructure Costing Estimator (DICE). The DICE method is novel in accounting for each country's demographic forecast trend, underlying population density, and future economic characteristics. The contribution of the paper to the literature is twofold. Firstly, to our knowledge the results present the first high-resolution global assessment which quantifies local universal broadband investment to achieve SDG Goal 9. Secondly, a high-resolution global assessment method is contributed which can inform future evaluations pertaining to universal broadband policy, providing insight on sub-national universal broadband investment. Previous studies focused only on using aggregated national data.

To provide universal broadband under baseline assumptions, the total investment needed is estimated to be $418 billion, or approximately 0.45 percent of global GDP (in 2020 US$). This estimate consists of $305 billion in emerging market economies (about 0.73 percent of their respective GDP) to serve approximately 2 billion unconnected citizens, and $102 billion in low-income developing countries (about 3.49 percent of their respective GDP) to connect 1.5 billion unconnected citizens. In advanced economies, additional investment needs are small and amount to $11 billion (about 0.02 percent of GDP) to connect 32 million currently unconnected citizens.

Sensitivity analysis of key model parameters demonstrates that future broadband policy assessment must be explicit about the quantity of data each user can consume. For example, when the data consumption rate is lowered to 10-20 GB/Month per user from a baseline of from 40-50 GB/Month per user, the total cost decreases by about half (to $201 bn from $418 bn), whereas raising the data consumption rate to 80-100 GB/Month per user results in a cost mark-up of up to 86 percent (to $783 bn from $418 bn).

The additional investments required to provide universal broadband largely depend on (i) the level of past historical investment in 4G and previous cellular generation technologies and (ii) the number of new potential users to be covered. In Sub-Saharan Africa, where there have been lower investments into newer generations of cellular technologies such as 4G, the Average Revenue Per User (ARPU) has been very low. This is compounded by many unconnected users (0.75 billion) and a large geographic region of 19.8 million square kilometers. The largest estimated investment is for Emerging and Developing Asia, equating to approximately $176 billion and reflecting the need to cover the vast population of 1.7 billion unconnected citizens across 19.5 million square kilometers.

These results focus on the specific investment costs in different income groups and regions based on business-as-usual approach to existing infrastructure deployment. There is a larger question as to whether this investment will actually take place in reality. However, there are certain options that can help in delivering universal broadband infrastructure. For example, technology, business model, regulatory, and policy changes can be carried out to help to (i) reduce the costs of infrastructure delivery and (ii) increase both the level of demand-side adoption (number of subscriptions) and the amount consumers can spend on mobile broadband (the ARPU per month). One assessment finds that if all cost-reducing measures are implemented, the necessary investment for providing universal broadband to all low- and middle-income countries could be reduced by half, avoiding the need for governments to provide public subsidies (Oughton et al., 2022). Importantly, as this assessment has not considered what percentage of investment will be delivered by the market vis-à-vis areas of market failure (which will require government support), decision makers must consider this key issue. Indeed, public subsidies or other governmental support should be targeted at areas where market actors cannot rationally deploy broadband infrastructure viably using market methods.

Finally, it is worth discussing and reflecting on three key limitations. Firstly, a major issue mediating investment in universal broadband infrastructure is the degree to which countries can deliver strong public investment management and capital spending implementation (Baum et al., 2020; Schwartz et al., 2020; Yoo et al., 2022), particularly in Sub-Saharan Africa (Lledó and Poplawski-Ribeiro, 2013).

Secondly, in relation to the methodological approach developed here, any global analysis will inevitably need to sacrifice country-specific detail out of necessity, in favor of providing systematic assessment of all countries. This primarily occurs due to the need to use globally harmonized datasets as model inputs, which are often more limited in nuance compared to country-specific data. As demonstrated here, while it is certainly very useful to systematically evaluate necessary investment across all countries, such activities are by no means a substitute for detailed country modeling, which is a necessary next step for identifying country-specific investment strategies for universal broadband infrastructure. A key shortcoming is the need to create simplifying data values and assumptions for country groups, which often averages away the unique circumstances of specific countries.

Finally, the techno-economic model developed provides decision-support insight based on a deductive scientific philosophy, contrasting with the inductive approach often adopted when utilizing observational statistical data, for example, in econometric assessments. In deductive research, an outcome can only be true if the premises set is true, providing a key limitation (although deductive research enables the testing of specific premises sets, which is the key advantage demonstrated in this paper).

There are numerous future research avenues. Firstly, it would be useful to explore the development of a more detailed demand-side model for each country which more broadly reflects demand constraints, such as affordability and digital literacy. Education is a key factor in overcoming the digital divide so users can see clear value in using connected devices, therefore more research is needed to properly quantify the investment needs to address this pertinent issue. Secondly, the approach adopted here is a feasible low-cost way to deliver universal broadband using, predominantly, wireless technologies, such as 4G and satellite broadband (due to their cost effectiveness in challenging deployment locations). However, beyond immediate needs, it is highly likely that the eventual goal over multiple decades will be deploying Fiber-To-The-Premises (FTTP). Therefore, future research should explore the cost implications of building FTTP infrastructure globally, on a country-by-country basis. Finally, as this assessment focused on the cost of building necessary broadband infrastructure, future research should explore the economic benefits of such investment.

# References

Abidi, N., El Herradi, M., Sakha, S., 2022. Digitalization and Resilience: Firm-level Evidence During the COVID-19 Pandemic.
Abrardi, L., Cambini, C., 2019. Ultra-fast broadband investment and adoption: A survey. Telecommunications Policy 43, 183–198. https://doi.org/10.1016/j.telpol.2019.02.005
Alliance for Affordable Internet, 2022. Affordable Internet is "1 for 2" [WWW Document]. Alliance for Affordable Internet. URL https://a4ai.org/affordable-internet-is-1-for-2/ (accessed 12.2.22).
Alper, E., Miktus, M., 2019a. Digital Connectivity in sub-Saharan Africa: A Comparative Perspective.
Alper, E., Miktus, M., 2019b. Bridging the Mobile Digital Divide in Sub-Saharan Africa: Costing under Demographic Change and Urbanization. IMF Working Papers, Working Paper No. 2019/249.
Baarsma, B., Groenewegen, J., 2021. COVID-19 and the Demand for Online Grocery Shopping: Empirical Evidence from the Netherlands. De Economist 169, 407–421. https://doi.org/10.1007/s10645-021-09389-y
Baum, A., Mogues, T., Verdier, G., 2020. Getting the most from public investment. Well Spent: How Strong Infrastructure Governance Can End Waste in Public Investment 30–49.
Bertschek, I., Briglauer, W., Hüschelrath, K., Kauf, B., Niebel, T., 2015. The Economic Impacts of Broadband Internet: A Survey. Review of Network Economics 14, 201–227. https://doi.org/10.1515/rne-2016-0032
Briglauer, W., Dürr, N., Gugler, K., 2021. A retrospective study on the regional benefits and spillover effects of high-speed broadband networks: Evidence from German counties. International Journal of Industrial Organization 74, 102677. https://doi.org/10.1016/j.ijindorg.2020.102677
Briglauer, W., Gugler, K., 2019. Go for Gigabit? First Evidence on Economic Benefits of High-speed Broadband Technologies in Europe. JCMS: Journal of Common Market Studies 57, 1071–1090. https://doi.org/10.1111/jcms.12872
Briglauer, W., Palan, N., 2023. Economic Benefits of High-Speed Broadband Coverage and Adoption, in: 5G, Artificial Intelligence, and Next Generation Internet of Things: Digital Innovation for Green and Sustainable Economies. IGI Global, pp. 27–43. https://doi.org/10.4018/978-1-6684-8634-4.ch002
Cave, M., 2018a. How disruptive is 5G? Telecommunications Policy, The implications of 5G networks: Paving the way for mobile innovation? 42, 653–658. https://doi.org/10.1016/j.telpol.2018.05.005
Cave, M., 2018b. How disruptive is 5G? Telecommunications Policy, The implications of 5G networks: Paving the way for mobile innovation? 42, 653–658. https://doi.org/10.1016/j.telpol.2018.05.005
Chang, H.-H., Meyerhoefer, C.D., 2021. COVID-19 and the Demand for Online Food Shopping Services: Empirical Evidence from Taiwan. American Journal of Agricultural Economics 103, 448–465. https://doi.org/10.1111/ajae.12170
Chen, Y., Oughton, E., Zagdanski, J., Jia, M., Tyler, P., 2022. Crowdsourced Data Indicates Broadband Has a Positive Impact on Local Business Creation. https://doi.org/10.2139/ssrn.4278740
Chinn, M.D., Fairlie, R.W., 2007. The determinants of the global digital divide: a cross-country analysis of computer and internet penetration. Oxford Economic Papers 59, 16–44. https://doi.org/10.1093/oep/gpl024
Copestake, A., Estefania-Flores, J., Furceri, D., 2022. Digitalization and Resilience. IMF.
Czernich, N., Falck, O., Kretschmer, T., Woessmann, L., 2011. Broadband infrastructure and economic growth. The Economic Journal 121, 505–532. https://doi.org/10.1111/j.1468-0297.2011.02420.x
del Portillo, I., Eiskowitz, S., Crawley, E.F., Cameron, B.G., 2021. Connecting the other half: Exploring options for the 50% of the population unconnected to the internet. Telecommunications Policy 45, 102092. https://doi.org/10.1016/j.telpol.2020.102092
DeStefano, T., Kneller, R., Timmis, J., 2018. Broadband infrastructure, ICT use and firm performance: Evidence for UK firms. Journal of Economic Behavior & Organization 155, 110–139. https://doi.org/10.1016/j.jebo.2018.08.020
Dunnewijk, T., Hultén, S., 2007. A brief history of mobile communication in Europe. Telematics and Informatics, Mobile Communications: From Cellular to Ad-hoc and Beyond 24, 164–179. https://doi.org/10.1016/j.tele.2007.01.013
Edquist, H., Goodridge, P., Haskel, J., Li, X., Lindquist, E., 2018. How important are mobile broadband networks for the global economic development? Information Economics and Policy 45, 16–29. https://doi.org/10.1016/j.infoecopol.2018.10.001
Ericsson, 2022. Ericsson mobility report 2022 - November. Ericsson, Stockholm, Sweden.
Falk, M., Hagsten, E., 2021. Impact of high-speed broadband access on local establishment dynamics. Telecommunications Policy 45, 102104. https://doi.org/10.1016/j.telpol.2021.102104
Falkner, M., Devetsikiotis, M., Lambadaris, I., 2000. An overview of pricing concepts for broadband IP networks. IEEE Communications Surveys Tutorials 3, 2–13. https://doi.org/10.1109/COMST.2000.5340798
Ford, G.S., Koutsky, T.M., 2005. Broadband and economic development: a municipal case study from florida. Review of Urban & Regional Development Studies 17, 216–229. https://doi.org/10.1111/j.1467-940X.2005.00107.x
Forge, S., Vu, K., 2020. Forming a 5G strategy for developing countries: A note for policy makers. Telecommunications Policy 44, 101975. https://doi.org/10.1016/j.telpol.2020.101975

Frias, Z., Mendo, L., Oughton, E.J., 2020. How Does Spectrum Affect Mobile Network Deployments? Empirical Analysis Using Crowdsourced Big Data. IEEE Access 8, 190812–190821. https://doi.org/10.1109/ACCESS.2020.3031963

GADM, 2019. Global Administrative Areas Database (Version 3.6) [WWW Document]. URL https://gadm.org/ (accessed 7.11.19).

Gallardo, R., Whitacre, B., Kumar, I., Upendram, S., 2020. Broadband metrics and job productivity: a look at county-level data. Ann Reg Sci. https://doi.org/10.1007/s00168-020-01015-0

Genakos, C., Valletti, T., Verboven, F., 2018. Evaluating market consolidation in mobile communications. Econ Policy 33, 45–100. https://doi.org/10.1093/epolic/eix020

Giordani, M., Polese, M., Laya, A., Bertin, E., Zorzi, M., 2021. 6G Drivers for B2B Market, in: Shaping Future 6G Networks. John Wiley & Sons, Ltd, pp. 9–22. https://doi.org/10.1002/9781119765554.ch2

Giordani, M., Polese, M., Mezzavilla, M., Rangan, S., Zorzi, M., 2020. Toward 6G Networks: Use Cases and Technologies. IEEE Communications Magazine 58, 55–61. https://doi.org/10.1109/MCOM.001.1900411

GSMA, 2019. The 5G guide: A reference for operators. GSMA, London.

GSMA Intelligence, 2023. The spend of an era: mobile capex to reach $1.5 trillion for 2023–2030 [WWW Document]. URL https://data.gsmaintelligence.com/research/research/research-2023/the-spend-of-an-era-mobile-capex-to-reach-1-5-trillion-for-2023-2030 (accessed 8.10.23).

Gupta, M.S., Keen, M.M., Shah, M.A., Verdier, M.G., 2017. Digital Revolutions in Public Finance. International Monetary Fund.

Hasbi, M., Dubus, A., 2020. Determinants of mobile broadband use in developing economies: Evidence from Sub-Saharan Africa. Telecommunications Policy 44, 101944. https://doi.org/10.1016/j.telpol.2020.101944

Hussain, A., Batool, I., Akbar, M., Nazir, M., 2021. Is ICT an enduring driver of economic growth? Evidence from South Asian economies. Telecommunications Policy 45, 102202. https://doi.org/10.1016/j.telpol.2021.102202

IMF, 2023. World Economic Outlook - Frequently Asked Questions [WWW Document]. IMF. URL https://www.imf.org/en/Publications/WEO/frequently-asked-questions (accessed 8.11.23).

IMF, 2022. Country Labor Costs, IMF Staff Calculations. World Economic Outlook.

IMF, 2021. Country Population Forecasts, IMF Staff Calculations. World Economic Outlook.

International Telecommunication Union, 2020. Connecting Humanity. International Telecommunication Union, Geneva, Switzerland.

Ioannou, N., Logothetis, V., Petre, K., Tselekounis, M., Chipouras, A., Katsianis, D., Varoutas, D., 2021. Network modeling approaches for calculating wholesale NGA prices: A full comparison based on the Greek fixed broadband market. Telecommunications Policy 45, 102184. https://doi.org/10.1016/j.telpol.2021.102184

Isley, C., Low, S.A., 2022. Broadband adoption and availability: Impacts on rural employment during COVID-19. Telecommunications Policy 46, 102310. https://doi.org/10.1016/j.telpol.2022.102310

ITU, 2022. ITU-D ICT Statistics [WWW Document]. ITU. URL https://www.itu.int:443/en/ITU-D/Statistics/Pages/stat/default.aspx (accessed 8.9.22).

Jung, J., López-Bazo, E., 2020. On the regional impact of broadband on productivity: The case of Brazil. Telecommunications Policy 44, 101826. https://doi.org/10.1016/j.telpol.2019.05.002

Karim, M.S., Nahar, S., Demirbag, M., 2022. Resource-Based Perspective on ICT Use and Firm Performance: A Meta-analysis Investigating the Moderating Role of Cross-Country ICT Development Status. Technological Forecasting and Social Change 179, 121626. https://doi.org/10.1016/j.techfore.2022.121626

Khera, P., Stephanie, M., Ng, Y., Ogawa, S., Sahay, R., 2022. Digital Financial Inclusion in Emerging and Developing Economies: A New Index.

Kongaut, C., Bohlin, E., 2014. Unbundling and infrastructure competition for broadband adoption: Implications for NGA regulation. Telecommunications Policy, Special issue on Moving Forward with Future Technologies: Opening a Platform for All 38, 760–770. https://doi.org/10.1016/j.telpol.2014.06.003

Koutroumpis, P., 2019. The economic impact of broadband: Evidence from OECD countries. Technological Forecasting and Social Change 148, 119719. https://doi.org/10.1016/j.techfore.2019.119719

Kumar, S.K.A., Oughton, E.J., 2023. Infrastructure Sharing Strategies for Wireless Broadband. IEEE Communications Magazine 1–7. https://doi.org/10.1109/MCOM.005.2200698

Lebrusán, I., Toutouh, J., 2020. Using Smart City Tools to Evaluate the Effectiveness of a Low Emissions Zone in Spain: Madrid Central. Smart Cities 3, 456–478. https://doi.org/10.3390/smartcities3020025

Lee, H., Jeong, S., Lee, K., 2021. Estimating the deployment costs of broadband universal service via fiber networks in Korea. Telecommunications Policy 45, 102105. https://doi.org/10.1016/j.telpol.2021.102105

Lehr, W., Queder, F., Haucap, J., 2021. 5G: A new future for Mobile Network Operators, or not? Telecommunications Policy 45, 102086. https://doi.org/10.1016/j.telpol.2020.102086

Lehtonen, O., 2020. Population grid-based assessment of the impact of broadband expansion on population development in rural areas. Telecommunications Policy 44, 102028. https://doi.org/10.1016/j.telpol.2020.102028

Liu, C., Wang, L., 2021. Who is left behind? Exploring the characteristics of China's broadband non-adopting families. Telecommunications Policy 45, 102187. https://doi.org/10.1016/j.telpol.2021.102187

Lledó, V., Poplawski-Ribeiro, M., 2013. Fiscal Policy Implementation in Sub-Saharan Africa. World Development 46, 79–91. https://doi.org/10.1016/j.worlddev.2013.01.030

Lobo, B.J., Alam, M.R., Whitacre, B., 2020. Broadband speed and unemployment rates: Data and measurement issues. Telecommunications Policy 44, 101829. https://doi.org/10.1016/j.telpol.2019.101829

Lythreatis, S., Singh, S.K., El-Kassar, A.-N., 2022. The digital divide: A review and future research agenda. Technological Forecasting and Social Change 175, 121359. https://doi.org/10.1016/j.techfore.2021.121359

Mac Domhnaill, C., Mohan, G., McCoy, S., 2021. Home broadband and student engagement during COVID-19 emergency remote teaching. Distance Education 42, 465–493. https://doi.org/10.1080/01587919.2021.1986372

Manlove, J., Whitacre, B., 2019. An evaluation of the Connected Nation broadband adoption program. Telecommunications Policy 43, 101809. https://doi.org/10.1016/j.telpol.2019.02.003

Mogensen, P., Na, W., Kovacs, I.Z., Frederiksen, F., Pokhariyal, A., Pedersen, K.I., Kolding, T., Hugl, K., Kuusela, M., 2007. LTE Capacity Compared to the Shannon Bound, in: 2007 IEEE 65th Vehicular Technology Conference - VTC2007-Spring. pp. 1234–1238. https://doi.org/10.1109/VETECS.2007.260

Ofcom, 2018. Final statement: Mobile call termination market review 2018-21 [WWW Document]. Ofcom. URL https://www.ofcom.org.uk/consultations-and-statements/category-1/mobile-call-termination-market-review (accessed 5.25.20).

OpenCelliD, 2022. OpenCelliD - Largest Open Database of Cell Towers & Geolocation - by Unwired Labs [WWW Document]. OpenCelliD - Largest Open Database of Cell Towers & Geolocation - by Unwired Labs. URL https://opencellid.org/#zoom=16&lat=37.77889&lon=-122.41942 (accessed 8.9.22).

Osoro, O.B., Oughton, E.J., Wilson, A.R., Rao, A., 2023. Sustainability assessment of Low Earth Orbit (LEO) satellite broadband mega-constellations. https://doi.org/10.48550/arXiv.2309.02338

Oughton, E., 2023. Digital Infrastructure Cost Estimator (DICE) [WWW Document]. URL https://github.com/edwardoughton/dice (accessed 8.15.23).

Oughton, E.J., 2022. Policy Options for Digital Infrastructure Strategies: A Simulation Model for Affordable Universal Broadband in Africa. Telematics and Informatics 101908. https://doi.org/10.1016/j.tele.2022.101908

Oughton, E.J., Comini, N., Foster, V., Hall, J.W., 2022. Policy choices can help keep 4G and 5G universal broadband affordable. Technological Forecasting and Social Change 176, 121409. https://doi.org/10.1016/j.techfore.2021.121409

Oughton, E.J., Frias, Z., van der Gaast, S., van der Berg, R., 2019. Assessing the capacity, coverage and cost of 5G infrastructure strategies: Analysis of the Netherlands. Telematics and Informatics 37, 50–69. https://doi.org/10.1016/j.tele.2019.01.003

Oughton, E.J., Jha, A., 2021. Supportive 5G Infrastructure Policies are Essential for Universal 6G: Assessment Using an Open-Source Techno-Economic Simulation Model Utilizing Remote Sensing. IEEE Access 9, 101924–101945. https://doi.org/10.1109/ACCESS.2021.3097627

Oughton, E.J., Lehr, W., 2022. Surveying 5G Techno-Economic Research to Inform the Evaluation of 6G Wireless Technologies. IEEE Access 10, 25237–25257. https://doi.org/10.1109/ACCESS.2022.3153046

Oughton, E.J., Lehr, W., Katsaros, K., Selinis, I., Bubley, D., Kusuma, J., 2021. Revisiting Wireless Internet Connectivity: 5G vs Wi-Fi 6. Telecommunications Policy 45, 102127. https://doi.org/10.1016/j.telpol.2021.102127

Oughton, E.J., Russell, T., Johnson, J., Yardim, C., Kusuma, J., 2020. itmlogic: The Irregular Terrain Model by Longley and Rice. Journal of Open Source Software 5, 2266. https://doi.org/10.21105/joss.02266

Oughton, E.J., Tran, M., Jones, C.B., Ebrahimy, R., 2016. Digital communications and information systems, in: Hall, J.W., Tran, M., Hickford, A.J., Nicholls, R.J. (Eds.), The Future of National Infrastructure: A System-of-Systems Approach. Cambridge University Press, Cambridge, pp. 181–202. https://doi.org/10.1017/CBO9781107588745.010

Prado, T.S., Bauer, J.M., 2021. Improving broadband policy design using market data: A general framework and an application to Brazil. Telecommunications Policy 45, 102111. https://doi.org/10.1016/j.telpol.2021.102111

Prieger, J.E., 2013a. The broadband digital divide and the economic benefits of mobile broadband for rural areas. Telecommunications Policy 37, 483–502. https://doi.org/10.1016/j.telpol.2012.11.003

Prieger, J.E., 2013b. The broadband digital divide and the economic benefits of mobile broadband for rural areas. Telecommunications Policy 37, 483–502. https://doi.org/10.1016/j.telpol.2012.11.003

Ramírez-Hassan, A., Carvajal-Rendón, D.A., 2021. Specification uncertainty in modeling internet adoption: A developing city case analysis. Utilities Policy 70, 101218. https://doi.org/10.1016/j.jup.2021.101218

Rendon, J., Ajibulu, A., Konstantinou, K., Bradford, J., Zimmermann, G., Droste, H., Canto, R., 2019. A business case for 5G mobile broadband in a dense urban area. Telecommunications Policy 43, 101813. https://doi.org/10.1016/j.telpol.2019.03.002

Rendon, J., Bradford, J., Ajibulu, A., Pearson, K., Konstantinou, K., Osman, H., Zimmermann, G., 2021. A business case for 5G services in an industrial sea port area. Telecommunications Policy 102264.

Rendon Schneir, J., Xiong, Y., 2016. A cost study of fixed broadband access networks for rural areas. Telecommunications Policy, Technological change and the provision, consumption and regulation of services: Papers from a European ITS regional conference 40, 755–773. https://doi.org/10.1016/j.telpol.2016.04.002

Rendon Schneir, J., Xiong, Y., 2014. Cost analysis of network sharing in FTTH/PONs. IEEE Communications Magazine 52, 126–134. https://doi.org/10.1109/MCOM.2014.6871680

Rosston, G.L., Wallsten, S.J., 2020. Increasing low-income broadband adoption through private incentives. Telecommunications Policy 44, 102020. https://doi.org/10.1016/j.telpol.2020.102020

Salemink, K., Strijker, D., Bosworth, G., 2017. Rural development in the digital age: A systematic literature review on unequal ICT availability, adoption, and use in rural areas. Journal of Rural Studies 54, 360–371. https://doi.org/10.1016/j.jrurstud.2015.09.001

Sawng, Y., Kim, P., Park, J., 2021. ICT investment and GDP growth: Causality analysis for the case of Korea. Telecommunications Policy 45, 102157. https://doi.org/10.1016/j.telpol.2021.102157

Schneir, J.R., Mölleryd, B.G., Oughton, E., Mas-Machuca, C., 2023. Guest Editorial: Techno-Economic Analysis of Telecommunications Systems. IEEE Communications Magazine 61, 22–23. https://doi.org/10.1109/MCOM.2023.10047852

Schwartz, M.G., Fouad, M.M., Hansen, M.T.S., Verdier, M.G., 2020. Well Spent: How Strong Infrastructure Governance Can End Waste in Public Investment. International Monetary Fund.

Simione, F.F., Li, Y., 2021. The Macroeconomic Impacts of Digitalization in Sub-Saharan Africa: Evidence from Submarine Cables.

Stamopoulos, D., Dimas, P., Tsakanikas, A., 2022. Exploring the structural effects of the ICT sector in the Greek economy: A quantitative approach based on input-output and network analysis. Telecommunications Policy 46, 102332. https://doi.org/10.1016/j.telpol.2022.102332

Strover, S., Riedl, M.J., Dickey, S., 2021. Scoping new policy frameworks for local and community broadband networks. Telecommunications Policy 102171. https://doi.org/10.1016/j.telpol.2021.102171

Sun, T., 2021. Digital Banking Support to Small Businesses amid COVID-19.

Szász, L., Bálint, C., Csíki, O., Nagy, B.Z., Rácz, B.-G., Csala, D., Harris, L.C., 2022. The impact of COVID-19 on the evolution of online retail: The pandemic as a window of opportunity. Journal of Retailing and Consumer Services 69, 103089. https://doi.org/10.1016/j.jretconser.2022.103089

Tatem, A.J., 2017. WorldPop, open data for spatial demography. Sci Data 4, 1–4. https://doi.org/10.1038/sdata.2017.4

Thompson, H.G., Garbacz, C., 2011. Economic impacts of mobile versus fixed broadband. Telecommunications Policy 35, 999–1009. https://doi.org/10.1016/j.telpol.2011.07.004

TowerXchange, 2019a. TowerXchange Europe Dossier 2019. TowerXchange, London.

TowerXchange, 2019b. TowerXchange MENA Dossier 2019. TowerXchange, London.

TowerXchange, 2019c. TowerXchange Americas Dossier 2019. TowerXchange, London.

TowerXchange, 2018a. TowerXchange Asia Dossier 2018. TowerXchange, London.

TowerXchange, 2018b. TowerXchange Africa Dossier 2018. TowerXchange, London.

TowerXchange, 2017. TowerXchange CALA Dossier 2017. TowerXchange, London.

Tranos, E., Mack, E.A., 2015. Broadband Provision and Knowledge-Intensive Firms: A Causal Relationship? Regional Studies 1–14. https://doi.org/10.1080/00343404.2014.965136

United Nations, 2019. The Sustainable Development Goals [WWW Document]. United Nations Sustainable Development. URL https://www.un.org/sustainabledevelopment/ (accessed 1.3.20).

Veedu, S.N.K., Mozaffari, M., Höglund, A., Yavuz, E.A., Tirronen, T., Bergman, J., Wang, Y.-P.E., 2022. Toward Smaller and Lower-Cost 5G Devices with Longer Battery Life: An Overview of 3GPP Release 17 RedCap. IEEE Communications Standards Magazine 6, 84–90. https://doi.org/10.1109/MCOMSTD.0001.2200029

Vu, K., Hanafizadeh, P., Bohlin, E., 2020. ICT as a driver of economic growth: A survey of the literature and directions for future research. Telecommunications Policy 44, 101922. https://doi.org/10.1016/j.telpol.2020.101922

West, J., Mace, M., 2010. Browsing as the killer app: Explaining the rapid success of Apple's iPhone. Telecommunications Policy 34, 270–286. https://doi.org/10.1016/j.telpol.2009.12.002

Whitacre, B., 2008. Factors influencing the temporal diffusion of broadband adoption: evidence from Oklahoma. Ann Reg Sci 42, 661–679. https://doi.org/10.1007/s00168-007-0178-7

Whitacre, B., Gallardo, R., Strover, S., 2014. Broadband's contribution to economic growth in rural areas: Moving towards a causal relationship. Telecommunications Policy 38, 1011–1023. https://doi.org/10.1016/j.telpol.2014.05.005

World Bank, 2022. Individuals using the Internet (% of population) [WWW Document]. World Bank Open Data. URL https://data.worldbank.org/indicator/IT.NET.USER.ZS (accessed 8.8.22).

WorldPop, 2019. WorldPop :: Population [WWW Document]. URL https://www.worldpop.org/project/categories?id=3 (accessed 1.2.20).

Yoo, C.S., Lambert, J., Pfenninger, T.P., 2022. Municipal fiber in the United States: A financial assessment. Telecommunications Policy 46, 102292. https://doi.org/10.1016/j.telpol.2021.102292

**Supplementary Evidence**

**<u>S1 The evolution of broadband connectivity</u>**

A new generation of cellular connectivity has been developed, standardized, and deployed on almost a decadal basis. Cellular connectivity began modestly with 1G systems in the 1980s, followed by the hugely successful 2G Global System for Mobile (GSM) in the 1990s, which saw a rapid global uptake of cellular devices enabling mobile voice calling and text messaging. Next, the first data-capable technologies were made available via 3G using the Universal Mobile Telecommunications System (UTMS) in the early 2000s, although the uptake of devices that used this technology was slower than expected.

More recently, 5G has introduced new capabilities that were not possible with 4G, including enhanced mobile broadband and ultra-reliable low-latency communication (Rendon et al., 2021, 2019). The newly delivered technical capabilities of 5G are aimed at enabling a range of new use cases, including Virtual Reality (VR) and Augmented Reality (AR) systems, across a range of industrial sectors (education, healthcare, transportation, etc.). Given the established trend, and with research and development underway for the next generation of cellular technologies, we should expect the deployment of 6G from 2028 onwards, targeting a range of new use cases (Giordani et al., 2021, 2020).

While 5G or 6G may seem enticing for expanding digital connectivity, 4G provides a more practical solution. For many developing countries, the technology has several limitations which make it unsuitable for providing universal broadband (Forge and Vu, 2020), particularly the higher cost of 5G devices which is prohibitive for many low-income users . Also, the incremental value of 5G compared to 4G is not necessarily clear yet, as the technology has primarily been used to provide enhanced mobile broadband—essentially 4G, but faster (Cave, 2018b; Lehr et al., 2021). Therefore, 5G technology is not necessarily deemed suitable as a main approach to achieve universal broadband access, especially as 5G can have poor viability in the hardest-to-reach locations, which are the prime targets for universal broadband.

Despite important gains over recent years in connecting unconnected Internet users, access to 4G, which enables low-cost wide-area access to the Internet, is far from universal. As of 2021, it is estimated that approximately 87 percent of the global population was 'covered' by a basic 4G signal from at least one mobile operator (Figure S1). Most regions of the world have high 4G population coverage, except Sub-Saharan Africa and the Middle East, where 2G voice and basic 3G data access can be more prevalent. However, adoption levels remain much lower and many more infrastructure assets (e.g., sites, fiber etc.) need to be built to provide the additional capacity necessary to serve higher adoption levels.

Figure S1. Population Coverage by Type of Mobile Network (percent)

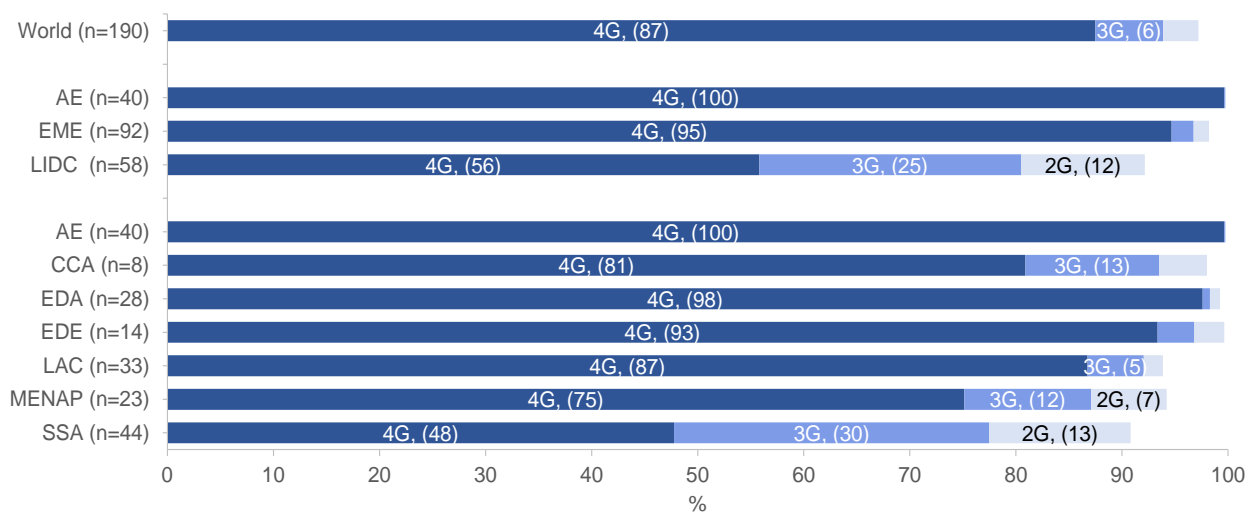

Source: ITU, IMF staff calculations.

Note: 'Coverage' is defined as the local population being served by at least one network operator. AE—Advanced Economies; EME—Emerging Market Economies; LIDC—Low Income Developing Countries; CCA—Caucasus and Central Asia; EDA—Emerging and Developing Asia; EDE—Emerging and Developing Europe; LAC—Latin America and the Caribbean; MENAP—Middle East, North Africa, Afghanistan, and Pakistan; SSA—Sub-Sahara Africa. The coverage percentage is weighted by population.

Within countries, there are also significant differences between urban, suburban, and rural settlements. The heterogeneity in access to the Internet and the availability of a computer is also quite considerable. Often technologies can be slow to reach ubiquity as they follow a logistic 's-shaped' adoption curve, with late adopters affected by a variety of barriers (Ramírez-Hassan and Carvajal-Rendón, 2021; Salemink et al., 2017; Whitacre, 2008). For example, adoption can be affected by challenges in supply-side infrastructure delivery and the stickiness of demand-side wages in low-income groups. In other words, the deployment of infrastructure gets harder and more complex, in economic viability terms, to serve the final unconnected population deciles.

The Alliance for Affordable Internet initially set an affordability threshold at whether 1 GB of mobile broadband data is priced at 2% or less of the average monthly income in a country. The UN Broadband Commission later adopted this "1 for 2" target as a standard for affordable Internet. More recently, this has been updated to 5 GB priced at less than 2% or less of the average monthly income in a country (Alliance for Affordable Internet, 2022). However, while 88% of the world's population live within areas covered by at least one 4G network (Figure S1), approximately one third of citizens remain offline due to the high cost of Internet access relative to income levels.

Data consumption per user varies greatly depending on the market. For example, in 2022 a mean global smartphone user consumed an average of 15 GB per month, but this is as low as 5 GB per month in Sub-Saharan Africa, or as high as 25 GB per month in India or parts of Southwest Asia (e.g., the UAE). Moreover, by 2028 this is forecast to reach 46 GB per month for a mean global smartphone user, compared to a low of 18 GB per month in Sub-Saharan Africa, and a high of 55 GB per month in North America and North East Asia (Ericsson, 2022).

COVID-19 has increased demand for Internet-based activities. The trends established before the pandemic were amplified as citizens turned to daily in-person tasks much more frequently via online options, such as working, shopping, financial management, education, and interacting with public services. For example, pre-pandemic (2018 and 2019) and during-the-pandemic household surveys in Brazil show that the demand for online services has increased (Figure S2).

Figure S2. Examples of Shifting Consumer Patterns Resulting from the COVID-19 Pandemic

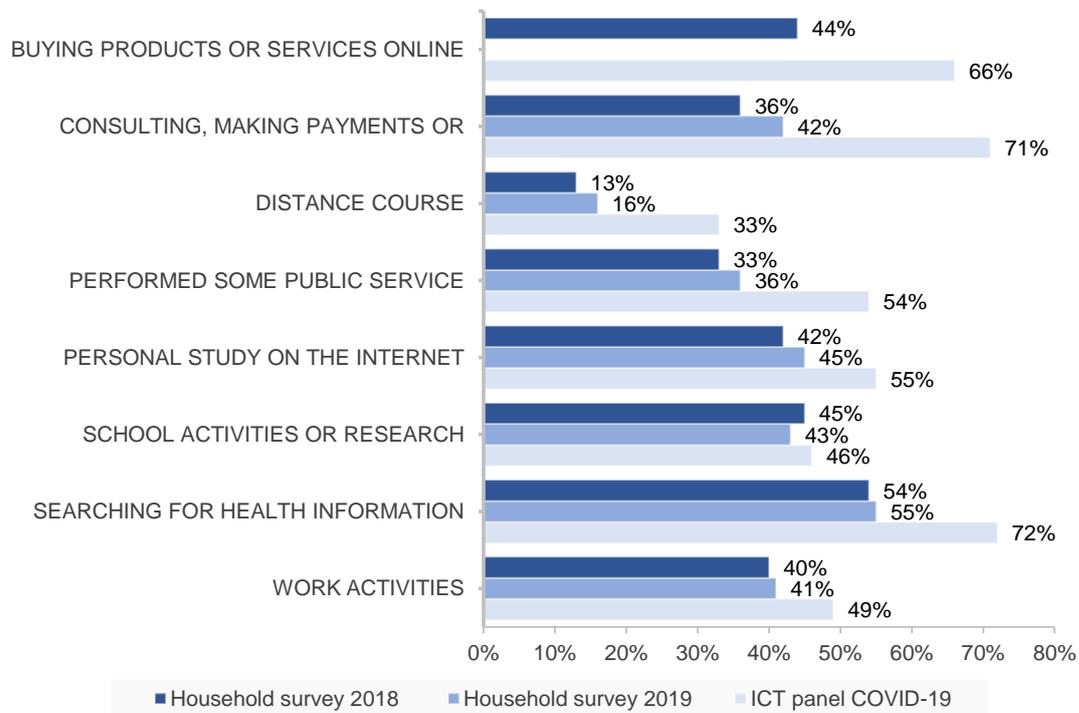

Source: CETIC Brazil and ITU (2020), https://www.itu.int/en/ITU-D/Statistics/Documents/facts/FactsFigures2020.pdf.

Note: Data for household survey 2018 and 2019 have a 12-month reference period, while the reference period for the COVID-19 panel was three months. Internet users aged 16 years and over.